
\font\ninerm=cmr9
\font\tenmmib=cmmib10
\newfam\bmitfam\textfont\bmitfam=\tenmmib
\def\bg{{\fam\bmitfam\mathchar"710D}}

\newread\epsffilein    
\newif\ifepsffileok    
\newif\ifepsfbbfound   
\newif\ifepsfverbose   
\newdimen\epsfxsize    
\newdimen\epsfysize    
\newdimen\epsftsize    
\newdimen\epsfrsize    
\newdimen\epsftmp      
\newdimen\pspoints     
\pspoints=1bp          
\epsfxsize=0pt         
\epsfysize=0pt         
\def\epsfbox#1{\global\def\epsfllx{72}\global\def\epsflly{72}%
   \global\def\epsfurx{540}\global\def\epsfury{720}%
   \def\lbracket{[}\def\testit{#1}\ifx\testit\lbracket
   \let\next=\epsfgetlitbb\else\let\next=\epsfnormal\fi\next{#1}}%
\def\epsfgetlitbb#1#2 #3 #4 #5]#6{\epsfgrab #2 #3 #4 #5 .\\%
   \epsfsetgraph{#6}}%
\def\epsfnormal#1{\epsfgetbb{#1}\epsfsetgraph{#1}}%
\def\epsfgetbb#1{%
%
%
\openin\epsffilein=#1
\ifeof\epsffilein\errmessage{I couldn't open #1, will ignore it}\else
%
%
   {\epsffileoktrue \chardef\other=12
    \def\do##1{\catcode`##1=\other}\dospecials \catcode`\ =10
    \loop
       \read\epsffilein to \epsffileline
       \ifeof\epsffilein\epsffileokfalse\else
%
%
          \expandafter\epsfaux\epsffileline:. \\%
       \fi
   \ifepsffileok\repeat
   \ifepsfbbfound\else
    \ifepsfverbose\message{No bounding box comment in #1; using defaults}\fi\fi
   }\closein\epsffilein\fi}%
%
%
\def\epsfclipstring{}
\def\epsfsetgraph#1{%
   \epsfrsize=\epsfury\pspoints
   \advance\epsfrsize by-\epsflly\pspoints
   \epsftsize=\epsfurx\pspoints
   \advance\epsftsize by-\epsfllx\pspoints
%
%
   \epsfxsize\epsfsize\epsftsize\epsfrsize
   \ifnum\epsfxsize=0 \ifnum\epsfysize=0
      \epsfxsize=\epsftsize \epsfysize=\epsfrsize
      \epsfrsize=0pt
%
%
     \else\epsftmp=\epsftsize \divide\epsftmp\epsfrsize
       \epsfxsize=\epsfysize \multiply\epsfxsize\epsftmp
       \multiply\epsftmp\epsfrsize \advance\epsftsize-\epsftmp
       \epsftmp=\epsfysize
       \loop \advance\epsftsize\epsftsize \divide\epsftmp 2
       \ifnum\epsftmp>0
          \ifnum\epsftsize<\epsfrsize\else
             \advance\epsftsize-\epsfrsize \advance\epsfxsize\epsftmp \fi
       \repeat
       \epsfrsize=0pt
     \fi
   \else \ifnum\epsfysize=0
     \epsftmp=\epsfrsize \divide\epsftmp\epsftsize
     \epsfysize=\epsfxsize \multiply\epsfysize\epsftmp
     \multiply\epsftmp\epsftsize \advance\epsfrsize-\epsftmp
     \epsftmp=\epsfxsize
     \loop \advance\epsfrsize\epsfrsize \divide\epsftmp 2
     \ifnum\epsftmp>0
        \ifnum\epsfrsize<\epsftsize\else
           \advance\epsfrsize-\epsftsize \advance\epsfysize\epsftmp \fi
     \repeat
     \epsfrsize=0pt
    \else
     \epsfrsize=\epsfysize
    \fi
   \fi
%
%
   \ifepsfverbose\message{#1: width=\the\epsfxsize, height=\the\epsfysize}\fi
   \epsftmp=10\epsfxsize \divide\epsftmp\pspoints
   \vbox to\epsfysize{\vfil\hbox to\epsfxsize{%
      \ifnum\epsfrsize=0\relax
        \includegraphics{#1}%
      \else
        \epsfrsize=10\epsfysize \divide\epsfrsize\pspoints
        \includegraphics{#1}%
      \fi
      \hfil}}%
\global\epsfxsize=0pt\global\epsfysize=0pt}%
%
%
{\catcode`\%=12 \global\let\epsfpercent=
%
%
\long\def\epsfaux#1#2:#3\\{\ifx#1\epsfpercent
   \def\testit{#2}\ifx\testit\epsfbblit
      \epsfgrab #3 . . . \\%
      \epsffileokfalse
      \global\epsfbbfoundtrue
   \fi\else\ifx#1\par\else\epsffileokfalse\fi\fi}%
%
%
\def\epsfempty{}%
\def\epsfgrab #1 #2 #3 #4 #5\\{%
\global\def\epsfllx{#1}\ifx\epsfllx\epsfempty
      \epsfgrab #2 #3 #4 #5 .\\\else
   \global\def\epsflly{#2}%
   \global\def\epsfurx{#3}\global\def\epsfury{#4}\fi}%
%
%
\def\epsfsize#1#2{\epsfxsize}
%
%

\newcount\Acount \Acount=0
\def\eqn#1{\global\advance\Acount by 1
  \global\xdef#1{\relax\the\Acount} \eqno{\dis#1}}
\def\dis#1{\hbox{(#1)}}
\def\spose#1{\hbox to 0pt{#1\hss}}
\def\simlt{\mathrel{\spose{\lower 3pt\hbox{$\mathchar"218$}}
     \raise 2.0pt\hbox{$\mathchar"13C$}}}
\def\simgt{\mathrel{\spose{\lower 3pt\hbox{$\mathchar"218$}}
     \raise 2.0pt\hbox{$\mathchar"13E$}}}

\def\ie{{\it i.e.}}
\def\etal{{\it et al.}}
\def\eg{{\it e.g.}}

\magnification=\magstep1
\font \authfont               = cmr10 scaled\magstep4
\font \fivesans               = cmss10 at 5pt
\font \headfont               = cmbx12 scaled\magstep4
\font \markfont               = cmr10 scaled\magstep1
\font \ninebf                 = cmbx9
\font \ninei                  = cmmi9
\font \nineit                 = cmti9
\font \ninerm                 = cmr9
\font \ninesans               = cmss10 at 9pt
\font \ninesl                 = cmsl9
\font \ninesy                 = cmsy9
\font \ninett                 = cmtt9
\font \sevensans              = cmss10 at 7pt
\font \sixbf                  = cmbx6
\font \sixi                   = cmmi6
\font \sixrm                  = cmr6
\font \sixsans                = cmss10 at 6pt
\font \sixsy                  = cmsy6
\font \smallescriptfont       = cmr5 at 7pt
\font \smallescriptscriptfont = cmr5
\font \smalletextfont         = cmr5 at 10pt
\font \subhfont               = cmr10 scaled\magstep4
\font \tafonts                = cmbx7  scaled\magstep2
\font \tafontss               = cmbx5  scaled\magstep2
\font \tafontt                = cmbx10 scaled\magstep2
\font \tams                   = cmmib10
\font \tamss                  = cmmib10
\font \tamt                   = cmmib10 scaled\magstep2
\font \tass                   = cmsy7  scaled\magstep2
\font \tasss                  = cmsy5  scaled\magstep2
\font \tast                   = cmsy10 scaled\magstep2
\font \tasys                  = cmex10 scaled\magstep1
\font \tasyt                  = cmex10 scaled\magstep2
\font \tbfonts                = cmbx7  scaled\magstep1
\font \tbfontss               = cmbx5  scaled\magstep1
\font \tbfontt                = cmbx10 scaled\magstep1
\font \tbms                   = cmmib10 scaled 833
\font \tbmss                  = cmmib10 scaled 600
\font \tbmt                   = cmmib10 scaled\magstep1
\font \tbss                   = cmsy7  scaled\magstep1
\font \tbsss                  = cmsy5  scaled\magstep1
\font \tbst                   = cmsy10 scaled\magstep1
\font \tenbfne                = cmb10
\font \tensans                = cmss10
\font \tpfonts                = cmbx7  scaled\magstep3
\font \tpfontss               = cmbx5  scaled\magstep3
\font \tpfontt                = cmbx10 scaled\magstep3
\font \tpmt                   = cmmib10 scaled\magstep3
\font \tpss                   = cmsy7  scaled\magstep3
\font \tpsss                  = cmsy5  scaled\magstep3
\font \tpst                   = cmsy10 scaled\magstep3
\font \tpsyt                  = cmex10 scaled\magstep3
\vsize=22.5true cm
\hsize=13.8true cm
\hfuzz=2pt
\tolerance=500
\abovedisplayskip=3 mm plus6pt minus 4pt
\belowdisplayskip=3 mm plus6pt minus 4pt
\abovedisplayshortskip=0mm plus6pt minus 2pt
\belowdisplayshortskip=2 mm plus4pt minus 4pt
\predisplaypenalty=0
\clubpenalty=10000
\widowpenalty=10000
\frenchspacing
\newdimen\oldparindent\oldparindent=1.5em
\parindent=1.5em
\skewchar\ninei='177 \skewchar\sixi='177
\skewchar\ninesy='60 \skewchar\sixsy='60
\hyphenchar\ninett=-1
\def\newline{\hfil\break}%
\catcode`@=11
\def\folio{\ifnum\pageno<\z@
\uppercase\expandafter{\romannumeral-\pageno}%
\else\number\pageno \fi}
\catcode`@=12 
  \mathchardef\Gamma="0100
  \mathchardef\Delta="0101
  \mathchardef\Theta="0102
  \mathchardef\Lambda="0103
  \mathchardef\Xi="0104
  \mathchardef\Pi="0105
  \mathchardef\Sigma="0106
  \mathchardef\Upsilon="0107
  \mathchardef\Phi="0108
  \mathchardef\Psi="0109
  \mathchardef\Omega="010A
  \mathchardef\bfGamma="0\the\bffam 00
  \mathchardef\bfDelta="0\the\bffam 01
  \mathchardef\bfTheta="0\the\bffam 02
  \mathchardef\bfLambda="0\the\bffam 03
  \mathchardef\bfXi="0\the\bffam 04
  \mathchardef\bfPi="0\the\bffam 05
  \mathchardef\bfSigma="0\the\bffam 06
  \mathchardef\bfUpsilon="0\the\bffam 07
  \mathchardef\bfPhi="0\the\bffam 08
  \mathchardef\bfPsi="0\the\bffam 09
  \mathchardef\bfOmega="0\the\bffam 0A

\def\sq{\hbox{\rlap{$\sqcap$}$\sqcup$}}

\def\utw{\smash{\rlap{\lower5pt\hbox{$\sim$}}}}
\def\udtw{\smash{\rlap{\lower6pt\hbox{$\approx$}}}}

\def\diameter{{\ifmmode\mathchoice
{\ooalign{\hfil\hbox{$\displaystyle/$}\hfil\crcr
{\hbox{$\displaystyle\mathchar"20D$}}}}
{\ooalign{\hfil\hbox{$\textstyle/$}\hfil\crcr
{\hbox{$\textstyle\mathchar"20D$}}}}
{\ooalign{\hfil\hbox{$\scriptstyle/$}\hfil\crcr
{\hbox{$\scriptstyle\mathchar"20D$}}}}
{\ooalign{\hfil\hbox{$\scriptscriptstyle/$}\hfil\crcr
{\hbox{$\scriptscriptstyle\mathchar"20D$}}}}
\else{\ooalign{\hfil/\hfil\crcr\mathhexbox20D}}%
\fi}}


\def\bbbc{{\mathchoice {\setbox0=\hbox{$\displaystyle\rm C$}\hbox{\hbox
to0pt{\kern0.4\wd0\vrule height0.9\ht0\hss}\box0}}
{\setbox0=\hbox{$\textstyle\rm C$}\hbox{\hbox
to0pt{\kern0.4\wd0\vrule height0.9\ht0\hss}\box0}}
{\setbox0=\hbox{$\scriptstyle\rm C$}\hbox{\hbox
to0pt{\kern0.4\wd0\vrule height0.9\ht0\hss}\box0}}
{\setbox0=\hbox{$\scriptscriptstyle\rm C$}\hbox{\hbox
to0pt{\kern0.4\wd0\vrule height0.9\ht0\hss}\box0}}}}
\def\bbbe{{\mathchoice {\setbox0=\hbox{\smalletextfont e}\hbox{\raise
0.1\ht0\hbox to0pt{\kern0.4\wd0\vrule width0.3pt height0.7\ht0\hss}\box0}}
{\setbox0=\hbox{\smalletextfont e}\hbox{\raise
0.1\ht0\hbox to0pt{\kern0.4\wd0\vrule width0.3pt height0.7\ht0\hss}\box0}}
{\setbox0=\hbox{\smallescriptfont e}\hbox{\raise
0.1\ht0\hbox to0pt{\kern0.5\wd0\vrule width0.2pt height0.7\ht0\hss}\box0}}
{\setbox0=\hbox{\smallescriptscriptfont e}\hbox{\raise
0.1\ht0\hbox to0pt{\kern0.4\wd0\vrule width0.2pt height0.7\ht0\hss}\box0}}}}
\def\bbbq{{\mathchoice {\setbox0=\hbox{$\displaystyle\rm Q$}\hbox{\raise
0.15\ht0\hbox to0pt{\kern0.4\wd0\vrule height0.8\ht0\hss}\box0}}
{\setbox0=\hbox{$\textstyle\rm Q$}\hbox{\raise
0.15\ht0\hbox to0pt{\kern0.4\wd0\vrule height0.8\ht0\hss}\box0}}
{\setbox0=\hbox{$\scriptstyle\rm Q$}\hbox{\raise
0.15\ht0\hbox to0pt{\kern0.4\wd0\vrule height0.7\ht0\hss}\box0}}
{\setbox0=\hbox{$\scriptscriptstyle\rm Q$}\hbox{\raise
0.15\ht0\hbox to0pt{\kern0.4\wd0\vrule height0.7\ht0\hss}\box0}}}}
\def\bbbt{{\mathchoice {\setbox0=\hbox{$\displaystyle\rm
T$}\hbox{\hbox to0pt{\kern0.3\wd0\vrule height0.9\ht0\hss}\box0}}
{\setbox0=\hbox{$\textstyle\rm T$}\hbox{\hbox
to0pt{\kern0.3\wd0\vrule height0.9\ht0\hss}\box0}}
{\setbox0=\hbox{$\scriptstyle\rm T$}\hbox{\hbox
to0pt{\kern0.3\wd0\vrule height0.9\ht0\hss}\box0}}
{\setbox0=\hbox{$\scriptscriptstyle\rm T$}\hbox{\hbox
to0pt{\kern0.3\wd0\vrule height0.9\ht0\hss}\box0}}}}
\def\bbbs{{\mathchoice
{\setbox0=\hbox{$\displaystyle     \rm S$}\hbox{\raise0.5\ht0\hbox
to0pt{\kern0.35\wd0\vrule height0.45\ht0\hss}\hbox
to0pt{\kern0.55\wd0\vrule height0.5\ht0\hss}\box0}}
{\setbox0=\hbox{$\textstyle        \rm S$}\hbox{\raise0.5\ht0\hbox
to0pt{\kern0.35\wd0\vrule height0.45\ht0\hss}\hbox
to0pt{\kern0.55\wd0\vrule height0.5\ht0\hss}\box0}}
{\setbox0=\hbox{$\scriptstyle      \rm S$}\hbox{\raise0.5\ht0\hbox
to0pt{\kern0.35\wd0\vrule height0.45\ht0\hss}\raise0.05\ht0\hbox
to0pt{\kern0.5\wd0\vrule height0.45\ht0\hss}\box0}}
{\setbox0=\hbox{$\scriptscriptstyle\rm S$}\hbox{\raise0.5\ht0\hbox
to0pt{\kern0.4\wd0\vrule height0.45\ht0\hss}\raise0.05\ht0\hbox
to0pt{\kern0.55\wd0\vrule height0.45\ht0\hss}\box0}}}}
\def\bbbz{{\mathchoice {\hbox{$\sans\textstyle Z\kern-0.4em Z$}}
{\hbox{$\sans\textstyle Z\kern-0.4em Z$}}
{\hbox{$\sans\scriptstyle Z\kern-0.3em Z$}}
{\hbox{$\sans\scriptscriptstyle Z\kern-0.2em Z$}}}}
\def\qed{\ifmmode\sq\else{\unskip\nobreak\hfil
\penalty50\hskip1em\null\nobreak\hfil\sq
\parfillskip=0pt\finalhyphendemerits=0\endgraf}\fi}
\newfam\sansfam
\textfont\sansfam=\tensans\scriptfont\sansfam=\sevensans
\scriptscriptfont\sansfam=\fivesans
\def\sans{\fam\sansfam\tensans}
\def\stackfigbox{\if
Y\FIG\global\setbox\figbox=\vbox{\unvbox\figbox\box1}%
\else\global\setbox\figbox=\vbox{\box1}\global\let\FIG=Y\fi}
\def\placefigure{\dimen0=\ht1\advance\dimen0by\dp1
\advance\dimen0by5\baselineskip
\advance\dimen0by0.4true cm
\ifdim\dimen0>\vsize\pageinsert\box1\vfill\endinsert
\else
\if Y\FIG\stackfigbox\else
\dimen0=\pagetotal\ifdim\dimen0<\pagegoal
\advance\dimen0by\ht1\advance\dimen0by\dp1\advance\dimen0by1.4true cm
\ifdim\dimen0>\pagegoal\stackfigbox
\else\box1\vskip4true mm\fi
\else\box1\vskip4true mm\fi\fi\fi}
%
\def\begfig#1cm#2\endfig{\par
\setbox1=\vbox{\dimen0=#1true cm\advance\dimen0
by1true cm\kern\dimen0#2}\placefigure}
\def\begdoublefig#1cm #2 #3 \enddoublefig{\begfig#1cm%
\vskip-.8333\baselineskip\line{\vtop{\hsize=0.46\hsize#2}\hfill
\vtop{\hsize=0.46\hsize#3}}\endfig}
\def\begfigsidebottom#1cm#2cm#3\endfigsidebottom{\dimen0=#2true cm
\ifdim\dimen0<0.4\hsize\message{Room for legend to narrow;
begfigsidebottom changed to begfig}\begfig#1cm#3\endfig\else
\par\def\figure##1##2{\vbox{\noindent\petit{\bf
Fig.\ts##1\unskip.\ }\ignorespaces ##2\par}}%
\dimen0=\hsize\advance\dimen0 by-.8true cm\advance\dimen0 by-#2true cm
\setbox1=\vbox{\hbox{\hbox to\dimen0{\vrule height#1true cm\hrulefill}%
\kern.8true cm\vbox{\hsize=#2true cm#3}}}\placefigure\fi}
\def\begfigsidetop#1cm#2cm#3\endfigsidetop{\dimen0=#2true cm
\ifdim\dimen0<0.4\hsize\message{Room for legend to narrow; begfigsidetop
changed to begfig}\begfig#1cm#3\endfig\else
\par\def\figure##1##2{\vbox{\noindent\petit{\bf
Fig.\ts##1\unskip.\ }\ignorespaces ##2\par}}%
\dimen0=\hsize\advance\dimen0 by-.8true cm\advance\dimen0 by-#2true cm
\setbox1=\vbox{\hbox{\hbox to\dimen0{\vrule height#1true cm\hrulefill}%
\kern.8true cm\vbox to#1true cm{\hsize=#2 true cm#3\vfill
}}}\placefigure\fi}
\def\figure#1#2{\vskip1true cm\setbox0=\vbox{\noindent\petit{\bf
Fig.\ts#1\unskip.\ }\ignorespaces #2\smallskip
\count255=0\global\advance\count255by\prevgraf}%
\ifnum\count255>1\box0\else
\centerline{\petit{\bf Fig.\ts#1\unskip.\
}\ignorespaces#2}\smallskip\fi}

\def\begtab#1cm#2\endtab{\par
   \ifvoid\topins\midinsert\medskip\vbox{#2\kern#1true cm}\endinsert
   \else\topinsert\vbox{#2\kern#1true cm}\endinsert\fi}
\def\begpet{\vskip6pt\bgroup\petit}
\def\endpet{\vskip6pt\egroup}
\newcount\frpages
\newcount\frpagegoal
\def\freepage#1{\global\frpagegoal=#1\relax\global\frpages=0\relax
\loop\global\advance\frpages by 1\relax
\message{Doing freepage \the\frpages\space of
\the\frpagegoal}\null\vfill\eject
\ifnum\frpagegoal>\frpages\repeat}
\newdimen\refindent
\def\begrefchapter#1{\titlea{}{\ignorespaces#1}%
\bgroup\petit
\setbox0=\hbox{1000.\enspace}\refindent=\wd0}
\def\ref{\goodbreak
\hangindent\oldparindent\hangafter=1
\noindent\ignorespaces}
\def\refno#1{\goodbreak
\hangindent\refindent\hangafter=1
\noindent\hbox to\refindent{#1\hss}\ignorespaces}
\def\endref{\goodbreak\endpet}
\def\vec#1{{\textfont1=\tams\scriptfont1=\tamss
\textfont0=\tenbf\scriptfont0=\sevenbf
\mathchoice{\hbox{$\displaystyle#1$}}{\hbox{$\textstyle#1$}}
{\hbox{$\scriptstyle#1$}}{\hbox{$\scriptscriptstyle#1$}}}}
\def\petit{\def\rm{\fam0\ninerm}%
\textfont0=\ninerm \scriptfont0=\sixrm \scriptscriptfont0=\fiverm
 \textfont1=\ninei \scriptfont1=\sixi \scriptscriptfont1=\fivei
 \textfont2=\ninesy \scriptfont2=\sixsy \scriptscriptfont2=\fivesy
 \def\it{\fam\itfam\nineit}%
 \textfont\itfam=\nineit
 \def\sl{\fam\slfam\ninesl}%
 \textfont\slfam=\ninesl
 \def\bf{\fam\bffam\ninebf}%
 \textfont\bffam=\ninebf \scriptfont\bffam=\sixbf
 \scriptscriptfont\bffam=\fivebf
 \def\sans{\fam\sansfam\ninesans}%
 \textfont\sansfam=\ninesans \scriptfont\sansfam=\sixsans
 \scriptscriptfont\sansfam=\fivesans
 \def\tt{\fam\ttfam\ninett}%
 \textfont\ttfam=\ninett
 \normalbaselineskip=11pt
 \setbox\strutbox=\hbox{\vrule height7pt depth2pt width0pt}%
 \normalbaselines\rm
\def\vec##1{{\textfont1=\tbms\scriptfont1=\tbmss
\textfont0=\ninebf\scriptfont0=\sixbf
\mathchoice{\hbox{$\displaystyle##1$}}{\hbox{$\textstyle##1$}}
{\hbox{$\scriptstyle##1$}}{\hbox{$\scriptscriptstyle##1$}}}}}
\nopagenumbers
%
\let\header=Y
\let\FIG=N
\newbox\figbox
\output={\if N\header\headline={\hfil}\fi\plainoutput\global\let\header=Y
\if Y\FIG\topinsert\unvbox\figbox\endinsert\global\let\FIG=N\fi}
\let\lasttitle=N
\def\bookauthor#1{\vfill\eject
     \bgroup
     \baselineskip=22pt
     \lineskip=0pt
     \pretolerance=10000
     \authfont
     \rightskip 0pt plus 6em
     \centerpar{#1}\vskip2true cm\egroup}
\def\bookhead#1#2{\bgroup
     \baselineskip=36pt
     \lineskip=0pt
     \pretolerance=10000
     \headfont
     \rightskip 0pt plus 6em
     \centerpar{#1}\vskip1true cm
     \baselineskip=22pt
     \subhfont\centerpar{#2}\vfill
     \parindent=0pt
     \baselineskip=16pt
     \leftskip=2.2true cm
     \markfont Springer-Verlag\newline
     Berlin Heidelberg New York\newline
     London Paris Tokyo Singapore\bigskip\bigskip
     [{\it This is page III of your manuscript and will be reset by
     Springer.}]
     \egroup\let\header=N\eject}
\def\centerpar#1{{\parfillskip=0pt
\rightskip=0pt plus 1fil
\leftskip=0pt plus 1fil
\advance\leftskip by\oldparindent
\advance\rightskip by\oldparindent
\def\newline{\break}%
\noindent\ignorespaces#1\par}}
\def\part#1#2{\vfill\supereject\let\header=N
\centerline{\subhfont#1}%
\vskip75pt
     \bgroup
\textfont0=\tpfontt \scriptfont0=\tpfonts \scriptscriptfont0=\tpfontss
\textfont1=\tpmt \scriptfont1=\tbmt \scriptscriptfont1=\tams
\textfont2=\tpst \scriptfont2=\tpss \scriptscriptfont2=\tpsss
\textfont3=\tpsyt \scriptfont3=\tasys \scriptscriptfont3=\tenex
     \baselineskip=20pt
     \lineskip=0pt
     \pretolerance=10000
     \tpfontt
     \centerpar{#2}
     \vfill\eject\egroup\ignorespaces}
\newtoks\AUTHOR
\newtoks\HEAD
\catcode`\@=\active
\def\author#1{\bgroup
\baselineskip=22pt
\lineskip=0pt
\pretolerance=10000
\markfont
\centerpar{#1}\bigskip\egroup
{\def@##1{}%
\setbox0=\hbox{\petit\kern2.5true cc\ignorespaces#1\unskip}%
\ifdim\wd0>\hsize
\message{The names of the authors exceed the headline, please use a }%
\message{short form with AUTHORRUNNING}\gdef\leftheadline{%
\hbox to2.5true cc{\folio\hfil}AUTHORS suppressed due to excessive
length\hfil}%
\global\AUTHOR={AUTHORS were to long}\else
\xdef\leftheadline{\hbox to2.5true
cc{\noexpand\folio\hfil}\ignorespaces#1\hfill}%
\global\AUTHOR={\def@##1{}\ignorespaces#1\unskip}\fi
}\let\INS=E}
\def\address#1{\bgroup
\centerpar{#1}\bigskip\egroup
\catcode`\@=12
\vskip2cm\noindent\ignorespaces}
\let\INS=N%
\def@#1{\if N\INS\unskip\ $^{#1}$\else\if
E\INS\noindent$^{#1}$\let\INS=Y\ignorespaces
\else\par
\noindent$^{#1}$\ignorespaces\fi\fi}%
\catcode`\@=12
\headline={\petit\def\newline{ }\def\fonote#1{}\ifodd\pageno
\rightheadline\else\leftheadline\fi}
\def\rightheadline{\hfil Missing CONTRIBUTION
title\hbox to2.5true cc{\hfil\folio}}
\def\leftheadline{\hbox to2.5true cc{\folio\hfil}Missing name(s) of the
author(s)\hfil}
\nopagenumbers
\let\header=Y

\let\lasttitle=N
 \def\contribution#1{\vfill\supereject
 \ifodd\pageno\else\null\vfill\supereject\fi
 \let\header=N\bgroup
 \textfont0=\tafontt \scriptfont0=\tafonts \scriptscriptfont0=\tafontss
 \textfont1=\tamt \scriptfont1=\tams \scriptscriptfont1=\tams
 \textfont2=\tast \scriptfont2=\tass \scriptscriptfont2=\tasss
 \par\baselineskip=16pt
     \lineskip=16pt
     \tafontt
     \raggedright
     \pretolerance=10000
     \noindent
     \centerpar{\ignorespaces#1}%
     \vskip12pt\egroup
     \nobreak
     \parindent=0pt
     \everypar={\global\parindent=1.5em
     \global\let\lasttitle=N\global\everypar={}}%
     \global\let\lasttitle=A%
     \setbox0=\hbox{\petit\def\newline{ }\def\fonote##1{}\kern2.5true
     cc\ignorespaces#1}\ifdim\wd0>\hsize
     \message{Your CONTRIBUTIONtitle exceeds the headline,
please use a short form
with CONTRIBUTIONRUNNING}\gdef\rightheadline{\hfil CONTRIBUTION title
suppressed due to excessive length\hbox to2.5true cc{\hfil\folio}}%
\global\HEAD={HEAD was to long}\else
\gdef\rightheadline{\hfill\ignorespaces#1\unskip\hbox to2.5true
cc{\hfil\folio}}\global\HEAD={\ignorespaces#1\unskip}\fi
\catcode`\@=\active
     \ignorespaces}
 \def\contributionnext#1{\vfill\supereject
 \let\header=N\bgroup
 \textfont0=\tafontt \scriptfont0=\tafonts \scriptscriptfont0=\tafontss
 \textfont1=\tamt \scriptfont1=\tams \scriptscriptfont1=\tams
 \textfont2=\tast \scriptfont2=\tass \scriptscriptfont2=\tasss
 \par\baselineskip=16pt
     \lineskip=16pt
     \tafontt
     \raggedright
     \pretolerance=10000
     \noindent
     \centerpar{\ignorespaces#1}%
     \vskip12pt\egroup
     \nobreak
     \parindent=0pt
     \everypar={\global\parindent=1.5em
     \global\let\lasttitle=N\global\everypar={}}%
     \global\let\lasttitle=A%
     \setbox0=\hbox{\petit\def\newline{ }\def\fonote##1{}\kern2.5true
     cc\ignorespaces#1}\ifdim\wd0>\hsize
     \message{Your CONTRIBUTIONtitle exceeds the headline,
please use a short form
with CONTRIBUTIONRUNNING}\gdef\rightheadline{\hfil CONTRIBUTION title
suppressed due to excessive length\hbox to2.5true cc{\hfil\folio}}%
\global\HEAD={HEAD was to long}\else
\gdef\rightheadline{\hfill\ignorespaces#1\unskip\hbox to2.5true
cc{\hfil\folio}}\global\HEAD={\ignorespaces#1\unskip}\fi
\catcode`\@=\active
     \ignorespaces}
\def\motto#1#2{\bgroup\petit\leftskip=6.5true cm\noindent\ignorespaces#1
\if!#2!\else\medskip\noindent\it\ignorespaces#2\fi\bigskip\egroup
\let\lasttitle=M
\parindent=0pt
\everypar={\global\parindent=\oldparindent
\global\let\lasttitle=N\global\everypar={}}%
\global\let\lasttitle=M%
\ignorespaces}
\def\abstract#1{\bgroup\petit\noindent
{\bf Abstract: }\ignorespaces#1\vskip28pt\egroup
\let\lasttitle=N
\parindent=0pt
\everypar={\global\parindent=\oldparindent
\global\let\lasttitle=N\global\everypar={}}%
\ignorespaces}
\def\titlea#1#2{\if N\lasttitle\else\vskip-28pt
     \fi
     \vskip18pt plus 4pt minus4pt
     \bgroup
\textfont0=\tbfontt \scriptfont0=\tbfonts \scriptscriptfont0=\tbfontss
\textfont1=\tbmt \scriptfont1=\tbms \scriptscriptfont1=\tbmss
\textfont2=\tbst \scriptfont2=\tbss \scriptscriptfont2=\tbsss
\textfont3=\tasys \scriptfont3=\tenex \scriptscriptfont3=\tenex
     \baselineskip=16pt
     \lineskip=0pt
     \pretolerance=10000
     \noindent
     \tbfontt
     \rightskip 0pt plus 6em
     \setbox0=\vbox{\vskip23pt\def\fonote##1{}%
     \noindent
     \if!#1!\ignorespaces#2
     \else\setbox0=\hbox{\ignorespaces#1\unskip\ }\hangindent=\wd0
     \hangafter=1\box0\ignorespaces#2\fi
     \vskip18pt}%
     \dimen0=\pagetotal\advance\dimen0 by-\pageshrink
     \ifdim\dimen0<\pagegoal
     \dimen0=\ht0\advance\dimen0 by\dp0\advance\dimen0 by
     3\normalbaselineskip
     \advance\dimen0 by\pagetotal
     \ifdim\dimen0>\pagegoal\eject\fi\fi
     \noindent
     \if!#1!\ignorespaces#2
     \else\setbox0=\hbox{\ignorespaces#1\unskip\ }\hangindent=\wd0
     \hangafter=1\box0\ignorespaces#2\fi
     \vskip18pt plus4pt minus4pt\egroup
     \nobreak
     \parindent=0pt
     \everypar={\global\parindent=\oldparindent
     \global\let\lasttitle=N\global\everypar={}}%
     \global\let\lasttitle=A%
     \ignorespaces}
 \def\titleb#1#2{\if N\lasttitle\else\vskip-28pt
     \fi
     \vskip18pt plus 4pt minus4pt
     \bgroup
\textfont0=\tenbf \scriptfont0=\sevenbf \scriptscriptfont0=\fivebf
\textfont1=\tams \scriptfont1=\tamss \scriptscriptfont1=\tbmss
     \lineskip=0pt
     \pretolerance=10000
     \noindent
     \bf
     \rightskip 0pt plus 6em
     \setbox0=\vbox{\vskip23pt\def\fonote##1{}%
     \noindent
     \if!#1!\ignorespaces#2
     \else\setbox0=\hbox{\ignorespaces#1\unskip\enspace}\hangindent=\wd0
     \hangafter=1\box0\ignorespaces#2\fi
     \vskip10pt}%
     \dimen0=\pagetotal\advance\dimen0 by-\pageshrink
     \ifdim\dimen0<\pagegoal
     \dimen0=\ht0\advance\dimen0 by\dp0\advance\dimen0 by
     3\normalbaselineskip
     \advance\dimen0 by\pagetotal
     \ifdim\dimen0>\pagegoal\eject\fi\fi
     \noindent
     \if!#1!\ignorespaces#2
     \else\setbox0=\hbox{\ignorespaces#1\unskip\enspace}\hangindent=\wd0
     \hangafter=1\box0\ignorespaces#2\fi
     \vskip8pt plus4pt minus4pt\egroup
     \nobreak
     \parindent=0pt
     \everypar={\global\parindent=\oldparindent
     \global\let\lasttitle=N\global\everypar={}}%
     \global\let\lasttitle=B%
     \ignorespaces}
 \def\titlec#1#2{\if N\lasttitle\else\vskip-23pt
     \fi
     \vskip18pt plus 4pt minus4pt
     \bgroup
\textfont0=\tenbfne \scriptfont0=\sevenbf \scriptscriptfont0=\fivebf
\textfont1=\tams \scriptfont1=\tamss \scriptscriptfont1=\tbmss
     \tenbfne
     \lineskip=0pt
     \pretolerance=10000
     \noindent
     \rightskip 0pt plus 6em
     \setbox0=\vbox{\vskip23pt\def\fonote##1{}%
     \noindent
     \if!#1!\ignorespaces#2
     \else\setbox0=\hbox{\ignorespaces#1\unskip\enspace}\hangindent=\wd0
     \hangafter=1\box0\ignorespaces#2\fi
     \vskip6pt}%
     \dimen0=\pagetotal\advance\dimen0 by-\pageshrink
     \ifdim\dimen0<\pagegoal
     \dimen0=\ht0\advance\dimen0 by\dp0\advance\dimen0 by
     2\normalbaselineskip
     \advance\dimen0 by\pagetotal
     \ifdim\dimen0>\pagegoal\eject\fi\fi
     \noindent
     \if!#1!\ignorespaces#2
     \else\setbox0=\hbox{\ignorespaces#1\unskip\enspace}\hangindent=\wd0
     \hangafter=1\box0\ignorespaces#2\fi
     \vskip6pt plus4pt minus4pt\egroup
     \nobreak
     \parindent=0pt
     \everypar={\global\parindent=\oldparindent
     \global\let\lasttitle=N\global\everypar={}}%
     \global\let\lasttitle=C%
     \ignorespaces}
 \def\titled#1{\if N\lasttitle\else\vskip-\baselineskip
     \fi
     \vskip12pt plus 4pt minus 4pt
     \bgroup
\textfont1=\tams \scriptfont1=\tamss \scriptscriptfont1=\tbmss
     \bf
     \noindent
     \ignorespaces#1\ \ignorespaces\egroup
     \ignorespaces}
\let\ts=\thinspace
\def\footnoterule{\kern-3pt\hrule width 2true cm\kern2.6pt}
\newcount\footcount \footcount=0
\def\advftncnt{\advance\footcount by1\global\footcount=\footcount}
\def\fonote#1{\advftncnt$^{\the\footcount}$\begingroup\petit
\parfillskip=0pt plus 1fil
\def\textindent##1{\hangindent0.5\oldparindent\noindent\hbox
to0.5\oldparindent{##1\hss}\ignorespaces}%
\vfootnote{$^{\the\footcount}$}{#1\vskip-9.69pt}\endgroup}
\def\item#1{\par\noindent
\hangindent6.5 mm\hangafter=0
\llap{#1\enspace}\ignorespaces}

\def\titleao#1{\vfill\supereject
\ifodd\pageno\else\null\vfill\supereject\fi
\let\header=N
     \bgroup
\textfont0=\tafontt \scriptfont0=\tafonts \scriptscriptfont0=\tafontss
\textfont1=\tamt \scriptfont1=\tams \scriptscriptfont1=\tamss
\textfont2=\tast \scriptfont2=\tass \scriptscriptfont2=\tasss
\textfont3=\tasyt \scriptfont3=\tasys \scriptscriptfont3=\tenex
     \baselineskip=18pt
     \lineskip=0pt
     \pretolerance=10000
     \tafontt
     \centerpar{#1}%
     \vskip75pt\egroup
     \nobreak
     \parindent=0pt
     \everypar={\global\parindent=\oldparindent
     \global\let\lasttitle=N\global\everypar={}}%
     \global\let\lasttitle=A%
     \ignorespaces}






\def\leaderfill{\kern0.5em\leaders\hbox to 0.5em{\hss.\hss}\hfill\kern
0.5em}
\newdimen\chapindent
\newdimen\sectindent
\newdimen\subsecindent
\newdimen\thousand
\setbox0=\hbox{\bf 10. }\chapindent=\wd0
\setbox0=\hbox{10.10 }\sectindent=\wd0
\setbox0=\hbox{10.10.1 }\subsecindent=\wd0
\setbox0=\hbox{\enspace 100}\thousand=\wd0
\def\contpart#1#2{\medskip\noindent
\vbox{\kern10pt\leftline{\textfont1=\tams
\scriptfont1=\tamss\scriptscriptfont1=\tbmss\bf
\advance\chapindent by\sectindent
\hbox to\chapindent{\ignorespaces#1\hss}\ignorespaces#2}\kern8pt}%
\let\lasttitle=Y\par}
\def\contcontribution#1#2{\if N\lasttitle\bigskip\fi
\let\lasttitle=N\line{{\textfont1=\tams
\scriptfont1=\tamss\scriptscriptfont1=\tbmss\bf#1}%
\if!#2!\hfill\else\leaderfill\hbox to\thousand{\hss#2}\fi}\par}
\def\conttitlea#1#2#3{\line{\hbox to
\chapindent{\strut\bf#1\hss}{\textfont1=\tams
\scriptfont1=\tamss\scriptscriptfont1=\tbmss\bf#2}%
\if!#3!\hfill\else\leaderfill\hbox to\thousand{\hss#3}\fi}\par}
\def\conttitleb#1#2#3{\line{\kern\chapindent\hbox
to\sectindent{\strut#1\hss}{#2}%
\if!#3!\hfill\else\leaderfill\hbox to\thousand{\hss#3}\fi}\par}
\def\conttitlec#1#2#3{\line{\kern\chapindent\kern\sectindent
\hbox to\subsecindent{\strut#1\hss}{#2}%
\if!#3!\hfill\else\leaderfill\hbox to\thousand{\hss#3}\fi}\par}
\long\def\lemma#1#2{\removelastskip\vskip\baselineskip\noindent{\tenbfne
Lemma\if!#1!\else\ #1\fi\ \ }{\it\ignorespaces#2}\vskip\baselineskip}
\long\def\proposition#1#2{\removelastskip\vskip\baselineskip\noindent{\tenbfne
Proposition\if!#1!\else\ #1\fi\ \ }{\it\ignorespaces#2}\vskip\baselineskip}
\long\def\theorem#1#2{\removelastskip\vskip\baselineskip\noindent{\tenbfne
Theorem\if!#1!\else\ #1\fi\ \ }{\it\ignorespaces#2}\vskip\baselineskip}
\long\def\corollary#1#2{\removelastskip\vskip\baselineskip\noindent{\tenbfne
Corollary\if!#1!\else\ #1\fi\ \ }{\it\ignorespaces#2}\vskip\baselineskip}
\long\def\example#1#2{\removelastskip\vskip\baselineskip\noindent{\tenbfne
Example\if!#1!\else\ #1\fi\ \ }\ignorespaces#2\vskip\baselineskip}
\long\def\exercise#1#2{\removelastskip\vskip\baselineskip\noindent{\tenbfne
Exercise\if!#1!\else\ #1\fi\ \ }\ignorespaces#2\vskip\baselineskip}
\long\def\problem#1#2{\removelastskip\vskip\baselineskip\noindent{\tenbfne
Problem\if!#1!\else\ #1\fi\ \ }\ignorespaces#2\vskip\baselineskip}
\long\def\solution#1#2{\removelastskip\vskip\baselineskip\noindent{\tenbfne
Solution\if!#1!\else\ #1\fi\ \ }\ignorespaces#2\vskip\baselineskip}


\long\def\definition#1#2{\removelastskip\vskip\baselineskip\noindent{\tenbfne
Definition\if!#1!\else\
#1\fi\ \ }\ignorespaces#2\vskip\baselineskip}
\def\frame#1{\bigskip\vbox{\hrule\hbox{\vrule\kern5pt
\vbox{\kern5pt\advance\hsize by-10.8pt
\centerline{\vbox{#1}}\kern5pt}\kern5pt\vrule}\hrule}\bigskip}
\def\frameddisplay#1#2{$$\vcenter{\hrule\hbox{\vrule\kern5pt
\vbox{\kern5pt\hbox{$\displaystyle#1$}%
\kern5pt}\kern5pt\vrule}\hrule}\eqno#2$$}
\def\typeset{\petit\noindent This book was processed by the author using
the \TeX\ macro package from Springer-Verlag.\par}
\outer\def\byebye{\bigskip\bigskip\typeset
\footcount=1\ifx\speciali\undefined\else
\loop\smallskip\noindent special character No\number\footcount:
\csname special\romannumeral\footcount\endcsname
\advance\footcount by 1\global\footcount=\footcount
\ifnum\footcount<11\repeat\fi
\gdef\leftheadline{\hbox to2.5true cc{\folio\hfil}\ignorespaces
\the\AUTHOR\unskip: \the\HEAD\hfill}\vfill\supereject\end}

\smallskip

\contribution{Concepts in
CMB Anisotropy Formation}
\author{Wayne Hu}
\address{Institute for Advanced Study, Princeton, NJ 08540}
\motto{``To divide is to leave something undivided.  To discriminate
between alternatives is to leave something which is neither
alternative.'' \smallskip\noindent
{\it Chuang-tzu}}
\abstract{
These lecture notes form a primer on the theory of cosmic
microwave background (CMB) anisotropy
formation.  With emphasis on conceptual aspects rather than
technical issues, we examine the physical foundations of
anisotropy evolution in relativistic kinetic and perturbation
theory as well as the manifestation of these principles in
primary and secondary anisotropies.  We discuss gauge choice
and gauge invariance and their use in understanding the
CMB.  Acoustic, gravitational
redshift and ionization effects have robust signatures
in the CMB spectrum and may allow determination of
classical cosmological parameters as well as reveal general distinctions
between models for structure formation.  We
develop the tight and weak coupling approximations as analytic tools
to help understand these effects and the robustness of their
signatures.
}

\vfill
\noindent whu@sns.ias.edu \hfill IASSNS-AST 95/52
\medskip
\hrule
\medskip
{\noindent\ninerm\baselineskip=12pt\hangindent=5ex\hangafter=1
To appear in {\sl The Universe at High-z, Large Scale Structure and
the Cosmic Microwave Background,} eds.
E. Martinez-Gonzalez and J.L Sanz (Springer Verlag, in press) \hfill}
\eject

\vphantom{mark}
\vskip 1 truein
\titlea{1}{Introduction}

More than three years has passed since cosmic microwave background
(CMB) anisotropies were first
detected at large angular scales in the {\it COBE} DMR
sky maps (Smoot \etal\ 1992).  Since then much progress has been made in
probing the anisotropy spectrum at smaller scales (see \eg\ the
compilations by Bond 1995; Scott, Silk \& White 1995; Steinhardt 1995).
Satellite missions
now being proposed would be able to map the anisotropy definitively
down to a fraction of a degree.  To exploit these recent and potential
experimental advances, we need an accurate and accessible theory
of anisotropy formation. The first part of this task was largely
accomplished in the 1980's with the ground-breaking work of
Wilson \& Silk (1981), Vittorio \& Silk (1984), and Bond \& Efstathiou
(1984).  Detailed numerical calculations in perturbation
theory allow accurate predictions of the anisotropy in most models
for structure formation.  Summaries of the current state of the
art in numerical codes can be found in Bond (1995), Hu \etal\ (1995)
and Ma \& Bertschinger (1995).

Numerical work however is notoriously inaccessible to
the uninitiated.  Moreover, it obscures the true potential of CMB
anisotropy spectrum for cosmology by limiting itself to specific
models.
Indeed, CMB anisotropies are sensitive to
classical cosmology parameters such as the matter and radiation
content, baryon fraction, expansion rate, curvature, and cosmological
constant (Bond \etal\ 1994; Seljak 1994; Hu \& Sugiyama 1995a) as well
as the model for structure formation (Hu \& Sugiyama 1995b; Albrecht
\etal\ 1995, Crittenden \& Turok 1995; Durrer \etal\ 1995; Magueijo
\etal\ 1995).  In these
lecture notes, we will review the basic concepts of anisotropy
formation and their consequence for cosmology.

We begin in \S 2
with relativistic
kinetic and perturbation theory which give the basic physical laws that
govern anisotropy formation.  Pure reductionism of this sort however
would miss the true use of CMB anisotropies.  In \S 3 and \S 4, we
map out the manifestations of the basic principles before
and after recombination, \ie\
the primary and secondary anisotropies.
These anisotropies possess information on
both the cosmological model and
structure formation.
{}From this decomposition and reconstruction,
the resulting complex structure of anisotropies can
be understood in a general model-independent manner.

\titlea{2}{Physical Foundation}

Two ingredients are necessary to describe the evolution of
cosmic microwave background anisotropies: relativistic
kinetic and perturbation theory. Kinetic theory describes
the radiation transport properties of the CMB photons in
the metric perturbed by gravitationally unstable density
fluctuations.  These
fluctuations are evolved through relativistic perturbation
theory.  Combined, the two considerations yield a complete
system that is responsible for anisotropy formation in any
model where structure formation proceeds by gravitational instability.

\titleb{2.1}{Relativistic Kinetic Theory}

Conceptually, relativistic kinetic theory is identical
to the familiar non-relativistic case: the phase space distribution
is conserved along geodesics save for a collision term due
to scattering,
$$
{d f \over dt} \equiv {\partial f \over \partial t} +
{\partial f \over \partial x^i}{ d x^i \over dt}
+ {\partial f \over \partial p}{ d p \over dt}
+ {\partial f \over \partial \gamma^i}{ d \gamma^i \over dt}
= C[f],
\eqn\eqnKinetic
$$
Here $\gamma_i$ are the direction cosines of the photon momentum $p$.
Before a redshift of $z_* \approx 1000$,
CMB photons were hot enough to ionize hydrogen.
Consequently, the dominant interaction process
for the CMB at early times was Compton scattering
off free electrons.  Due to the
higher electron density in the early universe, the Compton mean
free path was quite short, much smaller
than the particle horizon at that time.  The horizon
at $z_*$ subtends a degree or less for $\Omega_0 +\Omega_\Lambda
\simlt 1$. Thus on scales relevant for observable anisotropies, the
photons were tightly coupled to the electrons, which in turn
were tightly coupled to the protons by Coulomb interactions.
At $z_*$, neutral
hydrogen formed through ``recombination''
and the photons last scattered.  Unless the universe
suffered reionization at high redshift,
fluctuations in the CMB at recombination were frozen in
at $z_*$ and await observation today.

There are several features of Compton scattering worth noting
\medskip
\item {1.} Scattering couples the photons to the baryons and forces
perturbations in their number and hence energy density to evolve together.

\item {2.} Scattering isotropizes the photons in the electron rest frame
thus coupling the local CMB dipole to the electron velocity.

\item {3.} In the Thomson limit, there is no energy transfer in scattering.
Energy exchange only occurs to ${\cal O}(\left< v_e^2 \right>)$, \ie\
${\cal O}(T_e/m_e)$.

\item {4.} There is no change in photon number through Compton scattering.
\medskip
\noindent Scattering thus governs the intrinsic temperature
perturbations at last scattering and the dipole or
bulk velocity perturbation.
Since it does not change the net energy or photon number
in the CMB to lowest order, spectral distortions to the blackbody do not arise
in linear theory.  Unless the electrons have been heated significantly
above the temperature of the CMB or photons and/or energy has been
dumped into the CMB from an external source, the spectral information
can be ignored.

The remaining subtlety is that the photons propagate in a space-time
that is distorted by density fluctuations:
we must employ the geodesic equation in
the presence of perturbations.  This leads to gravitational redshift
effects from the $dp/dt$ term in equation \dis\eqnKinetic\
which can also generate fluctuations in the CMB.  The $d\gamma^i/dt$
term represents gravitational lensing and gives a first order
contribution only in a curved universe (see \S 3.4).
By integrating equation \dis\eqnKinetic\ over frequencies,
one obtains the {\it Boltzmann equation} for the evolution of
temperature perturbations
$\Theta(\eta,{\bf x},{\bg}) \equiv \Delta T/T$,
$$
\left[
{\partial \over \partial \eta} + \dot x^i {\partial \over \partial x^i}
+ \dot\gamma^i {\partial \over \partial \gamma^i}
\right] \Theta = S_G + S_C
\eqn\eqnBoltzImplicit
$$
where $S_G$ and $S_C$ are the sources from the gravitational
redshift and Compton scattering and overdots are derivatives with
respect to conformal time $\eta = \int dt/a$.
For a derivation from first principles
of the Compton collision term $S_C$,
see Hu, Scott \& Silk (1994), Dodelson \& Jubas (1995), and Kosowsky
(1995).
If the angular dependence and polarization of Compton scattering is
ignored
it reduces to $\dot \tau [ \Theta_0 -\Theta - \gamma_i v_e^i]$.
Here $\Theta_0$ is the isotropic temperature fluctuation and
the differential Compton optical depth $\dot \tau
 = x_e n_e \sigma_T a$ with $x_e$ as the ionization
fraction, $n_e$ as the electron density and $\sigma_T$ as
the Thomson cross section.
Implicit in $S_G$ and $S_C$ are the three fundamental
sources of anisotropies in the CMB:

\smallskip
\item{1.} Gravitational redshifts from the presence {\it and} evolution of
	  metric fluctuations.

\item{2.} Hot and cold spots from the intrinsic temperature
	  at last scattering.

\item{3.} The Doppler effect due to the velocity of the last scatterers.
\smallskip

\noindent
Of course, this high level description is not very practical.
Gravitational instability controls the evolution of
density, velocity, and metric perturbations in the universe.
All of
these
effects are thus related by relativistic perturbation theory.

\titleb{2.2}{Relativistic Perturbation Theory}

General relativity tells us that matter moves in a space-time
perturbed by fluctuations in the matter density itself.
Thus the Einstein equations reduce conceptually into two pieces.
The stress-energy tensor of the total matter is covariantly
conserved in the {\it perturbed} metric,
$T^{\mu\nu}_{\hphantom{\mu\nu};\mu}=0$, and matter
fluctuations are the source of metric perturbations via a generalized
Poisson equation.  For the former, $\nu=0$ gives number or energy density
conservation, \ie\ the continuity equation.  The spatial components
give momentum conservation, \ie\ the Euler equation.

\titlec{2.2.1}{Gauge Choice}

To explicitly implement these principles, we need to treat a
subtlety due to {\it gauge freedom} in general relativistic
perturbation theory.   In order to define a perturbation,
we must specify the relation between the
physical spacetime and the hypothetical unperturbed background.  The
difference between quantities at the {\it same coordinate values}
is deemed a perturbation. Since this choice is not unique,
one must generally fix a gauge before making any
calculations or interpretations regarding perturbation evolution.

\topinsert

\centerline{\hskip -0.75truecm \epsfxsize=4.0in \epsfbox{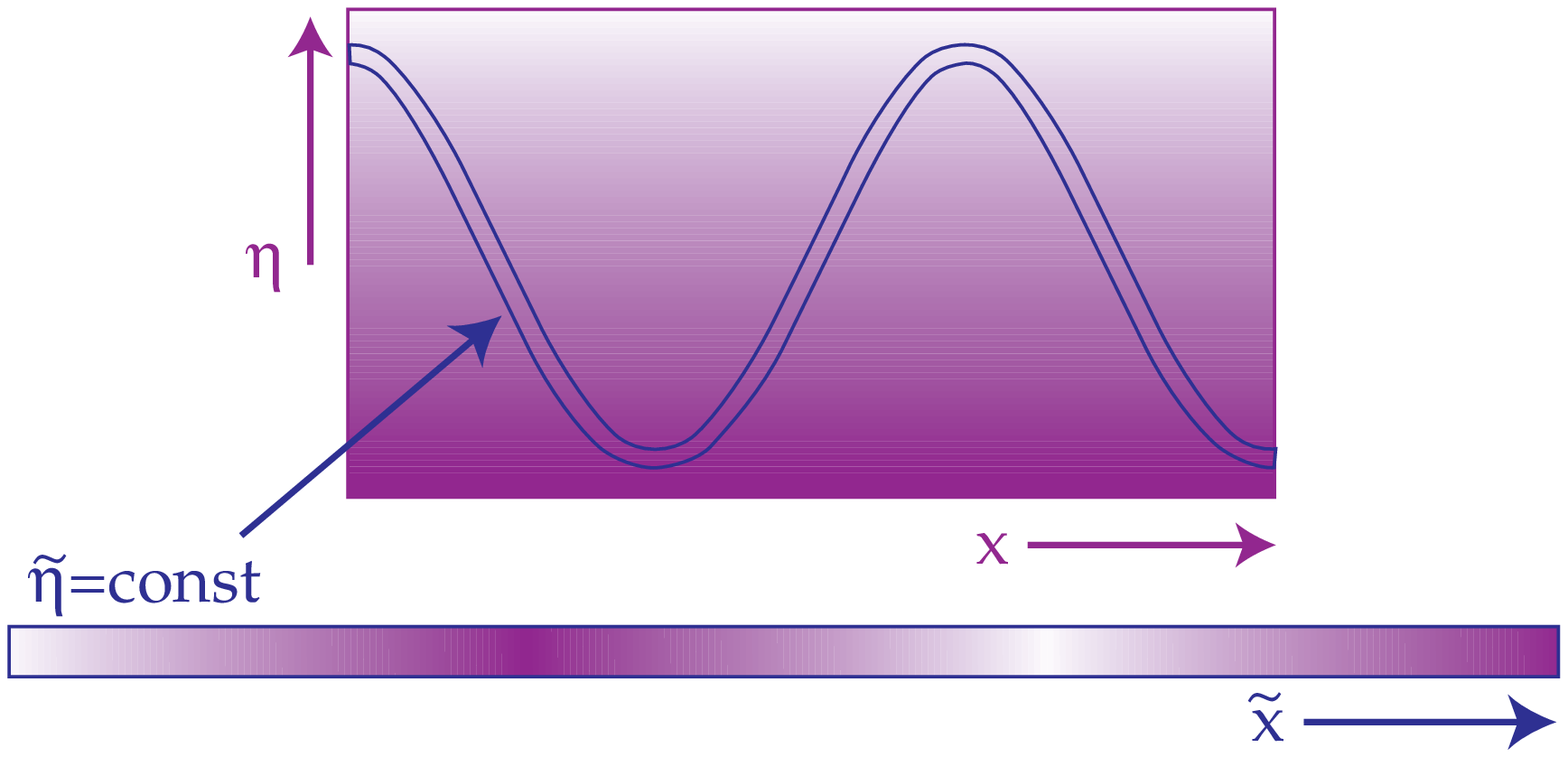}}

\vskip -0.75 truecm
\figure{1}{Gauge ambiguity.  Due to the change in density with
the expansion, even a homogeneous FRW universe can appear to have a
density perturbation if the time slicing is warped.  The synchronous
and total matter gauges choose the time slicing to correspond to
the rest frame of the freely falling and total matter respectively.
The Newtonian gauge chooses slicing such that the expansion rate
is shear free.}

\endinsert

\def\plane{e^{i {\bf k} \cdot {\bf x}}}
Practically speaking, a gauge choice involves fixing the constant-time
hypersurfaces and the spatial grid on these surfaces.
Suppose we warp the time slicing from some fiducial choice
by $\tilde \eta = \eta + T\plane$ (see Fig. 1).
The density of a fluid of particle $X$
on this slicing becomes
$\tilde \rho_X = \rho_X - \dot \rho_X T \plane$ for the same coordinate
value.  Because the expansion
makes densities decrease in time as
$$\dot \rho_X = -3(\rho_X+p_X){\dot a \over a},$$
in the new frame the density fluctuation is
$$
\tilde \delta_X = \delta_X + 3(1 + p_X/\rho_X){\dot a \over a}T \plane.
\eqn\eqnGaugeTrans
$$
As a simple example of gauge ambiguity, consider a
Friedmann-Robertson-Walker universe with {\it no} perturbations
$\delta_X =0$.
Even in this case, a density
fluctuation appears in the warped time slicing.
One might argue that this involves a
silly choice of time slicing and that observers may set up
a sensible coordinate system that removes ambiguities by exchanging
light signals.  This is true.  All ``sensible" choices of coordinate
systems agree on the nature and value of the density perturbation within
the horizon by requiring that the warping satisfy $\delta \eta/\eta = T/\eta
\ll \delta$ if $k \eta \gg 1$.
However outside the horizon, there is no ``sensible"
way to set up the coordinate system.  In particular, different
choices of time slicing can produce seemingly different results for
perturbations and their evolution in this regime.

There are two basic approaches to fixing a gauge or coordinate system.
We can choose a set of preferred observers and set the coordinate
frame to be their rest frame, \ie\ employ ``Lagrangian" coordinates.
The popular and useful {\it synchronous} gauge is
one such representation.
Here we choose freely falling observers to define the frame,
\eg\ the collisionless cold dark matter.  This rest frame
analysis is convenient
in that it simplifies the evolution equations and their solution.
An extension of this idea is employed in the {\it total matter} gauge.
Here the coordinate system coincides with the rest frame
of the combined relativistic and non-relativistic matter.  This
choice is computationally useful in the early universe where
radiation dominates and does not necessarily follow the perturbations
in the non-relativistic matter.

The computational convenience of these ``Lagrangian" techniques is
offset by difficulties in obtaining physical intuition for processes
such as anisotropy formation.  Perturbations grow by dilation
effects due to  distortions of
the coordinate grid
rather than simple causal mechanisms such as gravitational infall
and redshift.
An alternate choice is to take an ``Eulerian" coordinate system.
More specifically, instead of employing a set of preferred observers,
fix the coordinate system by a geometric property.  One useful choice
is the Newtonian gauge.  The coordinate frame here corresponds to
zero-shear hypersurfaces where the expansion appears isotropic.
Inside the horizon, this analysis reduces to the
standard non-relativistic Newtonian treatment.  Thus, our physical
intuition from the non-relativistic theory most easily carries
over to a complete treatment in this gauge.

Since there is a one-to-one mapping between gauges whose coordinates
are entirely fixed, these treatments are entirely equivalent and
give the same result for any {\it observable} quantity, \eg\
CMB anisotropies.  Ambiguities only
arise if the coordinate system is not entirely fixed.  In this
case, the one-to-many mapping produces unphysical {\it gauge modes}
which historically has produced much confusion (see Press \&
Vishniac 1980).  Gauge modes can
be avoided by completely fixing the coordinates.  The definitions
of the total matter and Newtonian gauges already do so.
On the other hand,
a synchronous coordinate system, defined by
a set of freely falling observers, requires us to
specify {\it which} set through the initial conditions.
In particular, if the initial velocity of the observers is fixed
then there are {\it no} gauge modes.  In practice, one usually
takes the rest frame of the CDM.

\titleb{2.2.2}{Gauge Invariance}

Due to the computational convenience of the ``Lagrangian" gauges
and the intuitive nature of the ``Eulerian" ones, it is often
useful to switch between gauges in the midst of a calculation
or even employ different gauge representations in a single
set of evolution equations.  A useful technique for
this purpose is provided by the {\it gauge invariant} method
(Bardeen 1980; Kodama \& Sasaki 1984; Mukhanov, Feldman \&
Brandenberger 1992).
It is called gauge invariant {\it not} because a density perturbation,
say, remains a density perturbation under a gauge transformation.
Rather, a density perturbation in a {\it specific}
gauge is represented in a coordinate free manner, \ie\ in a manner
valid in an arbitrary gauge.  In practical
terms, this just means that whatever gauge is chosen, one has
a systematic way to mix in quantities that represent perturbations
in a different gauge.

Let us make these considerations more concrete.  The most general
form of a metric perturbed by scalar fluctuations
a given Fourier mode $k$ is (Bardeen 1980)
\def\plane{e^{i {\bf k} \cdot {\bf x}}}
$$
\eqalign{
g_{00} & =  -a^2 [1 + 2 A^G \plane],  \cr
g_{0j} & =   a^2 B^G i \hat k_j \plane,  \cr
g_{ij} & =  a^2 \{ \delta_{ij} + [2 H_L^G \delta_{ij} + 2 H_T^G
		(-\hat k_i \hat k_j + \delta_{ij}/3)]\plane\}, \cr
}
\eqn\eqnMetric
$$
where for illustration purposes, we have taken a flat geometry and
superscript $G$ is meant to remind the reader that these quantities
are gauge dependent. By
fixing the gauge, we can eliminate two out of the four metric terms.
In the Newtonian gauge, $B^N = H_T^N = 0$, and it is traditional
to label the Newtonian potential $A^N = \Psi$ and the space
curvature perturbation $H_L^N = \Phi$.  Synchronous gauge
requires $A^S = B^S = 0$, whereas the total matter gauge takes
$B^T = V_T^T$ and $H_T^T=0$ where $V_T^T$ is the velocity of the
total matter.   Ordinarily, one now writes down the Einstein
equations $G_{\mu\nu} = 8\pi G T_{\mu\nu}$ which contain
the conservation $ T^{\mu\nu}_{\hphantom{\mu\nu};\mu}=0$
and ``Poisson'' equations, with
a specific choice of gauge.  For a fluid of particle
$X$, the stress energy momentum tensor is
$$
\eqalign{
T^0_{\hphantom{0}0} & = -(1 + \delta_X^G \plane), \cr
T^0_{\hphantom{0}i} & = (\rho_X + p_X)V_X^G (-i \hat k_i) \plane, \cr
T^i_{\hphantom{0}j} & = (p_X + \delta p_X^G \plane)
\delta^i_{\hphantom{i}j} + p_X \Pi_X
(-\hat k^i \hat k_j + \delta^i_{\hphantom{i}j}) \plane, \cr }
\eqn\eqnStressEnergy
$$
where $\Pi_X$ is the anisotropic stress of the fluid.

As an example of the gauge invariant program, let us
write the total matter gauge density perturbation in a
gauge-independent way.
The total matter gauge condition is
satisfied by shifting the time slicing from
an arbitrary gauge by $T = (V_T^G - B^G)/k$
[Kodama \& Sasaki 1984, eq.~(3.3a); Hu 1995, eq.~(4.88)].
Equation  \dis\eqnGaugeTrans\ tells us that the density perturbation
transforms as
$$
\delta^T_X = \delta^G_X + 3(1 + p_X/\rho_X) {\dot a \over a} (V_T^G -
		B^G)/k.
\eqn\eqnTotalExample
$$
In an arbitrary gauge $G$, this represents what the density
perturbation would be in the total matter rest frame.
This ``gauge invariant" definition can be used to employ total matter
gauge variables in a Newtonian or synchronous gauge treatment:
$$
\eqalign{
\delta^T_X & = \delta^N_X + 3{\dot a \over a} (1+ p_X/\rho_X)V^N_T/k \cr
           & = \delta^S_X + 3{\dot a \over a} (1+ p_X/\rho_X)V^S_T/k. \cr
}
\eqn\eqnTotal
$$
If fluctuations are {\it adiabatic}, the number
density fluctuations
of the matter and radiation evolve together, \ie\ their bulk velocities
are equal.  In this case, $V^S_T$ is equal to the freely-falling cold
dark matter velocity, commonly defined to be zero in synchronous
gauge.
Equation \dis\eqnTotal\ then tells us that
the total matter gauge and synchronous
gauge density perturbation are numerically equivalent.  In a matter
dominated (pressureless) universe, there is no fundamental scale
in these rest frame evolution equations and density fluctuations
evolve in a scale free manner.  This is not true for the Newtonian gauge.
Here fluctuations grow by the gravitational infall of the matter into
potential wells.  Due to causality,
this introduces a fundamental scale, the {\it horizon}
scale, into the evolution equations.  For example in the case of
adiabatic fluctuations, density perturbations are constant outside
the horizon and only grow (or decay) after the horizon has grown
larger than the wavelength.

Now let us see how gauge modes creep in.  The mapping of
synchronous perturbations onto any gauge with fixed coordinates
is unambiguous as we have seen.  However, since it is a many-to-one
operation, its inverse leaves additional gauge freedom.  Notice
from equation \dis\eqnTotal, knowledge of the total matter
or Newtonian gauge perturbations only fixes a certain combination of
density and velocity perturbations in synchronous gauge.
To remove the ambiguity, one must fix the initial synchronous velocity.
Since an initial velocity decays with the expansion as
$a^{-1}$, equation \dis\eqnTotal\ tells us that
a different choice will alter the behavior of
densities by an additional term proportional to $\dot a / a^2$
or $a^{-2}$ during radiation domination and $a^{-3/2}$ during
matter domination.  This is a {\it gauge mode}.

In summary, the ``gauge invariant" approach does
nothing to solve the gauge
ambiguity; no solution is necessary since gauge choice poses no
fundamental problems.  Yet even
though there is nothing particularly deep about the ``gauge invariant"
program, it is often useful.  It allows us to borrow Newtonian
and rest frame concepts for use in any gauge. In addition, by providing
a systematic way of mapping perturbations in a specific gauge
onto an arbitrary gauge, one automatically determines whether
the coordinates have been completely
fixed, \ie\ whether gauge modes have been entirely eliminated.

\titleb{2.3}{Newtonian Gauge Equations}

Now we are ready to state the evolution equations in an explicit form.
To avoid unnecessary confusion, we will
stay in the pure Newtonian gauge throughout this treatment.
We will hereafter drop the superscript $N$ with the understanding
that all perturbation variables are in the Newtonian gauge unless
otherwise specified.
The continuity and Euler equations for the photon
temperature in flat space are
$$
\eqalign{
\dot \Theta_0 & = - {k \over 3}\Theta_1 - \dot \Phi, \cr
\dot \Theta_1 & = k[\Theta_0 + \Psi - {1 \over 6} \Pi_\gamma ]
			- \dot\tau (\Theta_1 - V_b),\cr
}
\eqn\eqnPhoton
$$
where the term proportional to the Compton
differential optical depth $\dot\tau$
comes from momentum conservation in the scattering.
Recall that
$\Theta_0 = {1 \over 4}\delta_\gamma$ is the isotropic temperature
fluctuation, and $\Theta_1 = V_\gamma$ is the amplitude of the
photon dipole or
bulk velocity.
One can see that
when the optical depth to scattering is high, the photons become
isotropic in the electron-baryon rest frame, \ie\ the dipole moment
$\Theta_1 = V_b$.  The anisotropic stress of the
photons $\Pi_\gamma$ is directly proportional to its quadrupole
moment.  Since scattering makes the photons isotropic in the
baryon rest frame, the photon anisotropic stress is negligible
before recombination.  These equations can also be directly
obtained from kinetic theory.  In fact, they are contained in
the Legendre decomposition of
the Boltzmann equation \dis\eqnBoltzImplicit\ in Fourier space
which also gives the evolution of $\Pi_\gamma$ through
a hierarchy of higher angular moments (Wilson \& Silk 1981).

Aside from the usual velocity divergence source in the
continuity equation, there is a term dependent on the metric.
This is due to the gravitational redshift effects of time dilation.
As
the form of the metric $g_{ij} = -a^2 \delta_{ij}(1 + 2\Phi \plane)$
implies, it is entirely analogous to the cosmological redshift.
Heuristically, the presence of
matter curves or stretches space taking the wavelength of
the photon with it.  In the Euler equation, the Newtonian
potential $\Psi$ acts as a source of the dipole.
Gradients in the potential also induce gravitational blue and
red shifts as the photons fall into and climb out of potential
wells ($\Psi < 0$).  This is countered by photon pressure from $\Theta_0$.
As the temperature rises so does the pressure which opposes
the fall of a photon into the potential well.  As the photons
free stream, power in the
dipole is converted into the higher multipole moments through
the quadrupole
via the $\Pi_\gamma$ term.

The baryon continuity and Euler equations take on a similar
form
$$
\eqalign{
\dot \delta_b & = - k V_b -  3 \dot \Phi,  \cr
\dot V_b & = - {\dot a \over a} V_b + k\Psi + \dot\tau(\Theta_1
	- V_b)/R.
}
\eqn\eqnBaryon
$$
Momentum conservation in Compton scattering gives the form
of the coupling.  From equation \dis\eqnStressEnergy,
the effective momentum density of a general
fluid $X$ is $(\rho_X + p_X)V_X$.  Since $p_\gamma = {1 \over 3}
\rho_\gamma$ and $p_b \ll \rho_b$, conservation implies
$
(4\rho_\gamma/3) \delta \Theta_1 = \rho_b \delta V_b.
$
Thus the Compton coupling for the baryons takes on a similar
form to the protons but is of opposite sign and is altered by a
factor of $R=3\rho_b/4\rho_\gamma$.

Again, the continuity equation is modified by gravitational effects
from the stretching of space associated with $\Phi$.  Since the
density decreases as the length scale cubed, the total differential
effect becomes $3\dot\Phi$.  In the Euler equation, the velocity is
damped by the expansion and enhanced by infall into potential wells.
Particle momenta scale as $a^{-1}$ due to the expansion.  For
non-relativistic particles, this causes the peculiar velocity to
scale similarly in the absence of sources.  In the fully-relativistic
case, it causes the temperature to decrease as $a^{-1}$.

Decoupled components
such as the massless neutrinos and cold dark matter follow identical
evolution equations save for the absence of Compton coupling.
The metric perturbations on the other hand feel the influence of the total
matter perturbations,
$\rho_T \delta_T  = \sum_i \rho_i \delta_i,$
$(\rho_T + p_T) V_T  = \sum_i (\rho_i + p_i)V_i,$ and
$p_T \Pi_T = \sum_i p_i \Pi_i, $
where $i$ runs through all the particle species,
through the generalized Poisson equations,
$$
\eqalign{
k^2 \Phi & = 4\pi G a^2 \rho_T [\delta_T + 3{\dot a \over a}
		(1 + p_T/\rho_T) V_T/k], \cr
k^2 (\Psi + \Phi) & = -8\pi G a^2 p_T \Pi_T, \cr
}
\eqn\eqnPoisson
$$
which arise from the time-time $+$ time-space and traceless space-space
components of the Einstein equations respectively.
The presence of $a^2$ in the first equation represents the conversion
from physical to comoving coordinates.
Notice that if we rewrite the equation in terms of the density fluctuation
on the total matter rest frame, the first equation would
take on the familiar non-relativistic form of the Poisson equation
and simplify perturbation calculations.  The additional term
represents a relativistic effect that is important outside the
horizon.  When anisotropic stress $p_T \Pi_T$
may be ignored, the second
of equations \dis\eqnPoisson\ reduces to $\Psi = -\Phi$ as one would
expect for the Newtonian potential.

\titleb{2.4}{Gauge Tricks: Sachs-Wolfe Example}

Examining the photon conservation equations \dis\eqnPhoton\ and the
Poisson equations \dis\eqnPoisson, we see that the former are
simpler in the Newtonian gauge, whereas the latter are simpler in
the total matter gauge.  Let us see how gauge tricks developed

in \S 2.2 can help us understand their joint evolution.
In the matter dominated epoch, the Poisson and continuity
equations reduce to
$$
\Phi = {6 \over (k\eta)^2} \delta_T^T, \qquad
\dot \delta_T^T = -k V_T,
$$
as one expects from a non-relativistic analysis.  Since $\delta_T^T$
is the density perturbation on the matter rest frame, its evolution
in the growing mode also satisfies the non-relativistic relation
$\delta_T^T \propto \eta^2$. Hence $V_T = -2 \delta_T^T/k\eta$.  Recalling its
relation to
the Newtonian density perturbation, we obtain
$$
\eqalign{
\delta_T & = \delta_T^T - {6 \over k\eta}V_T = \left[ 1 + {12 \over
 (k\eta)^2} \right] \delta_T^T \cr
	 & \approx {12 \over (k\eta)^2} \delta_T^T = 2\Phi,
\qquad k\eta \ll 1.\cr}
$$
For adiabatic fluctuations, the intrinsic temperature fluctuation,
$\Theta_0 \equiv {1 \over 4}\delta_\gamma
= {1 \over 3}\delta_T = {2 \over 3}\Phi$.  We shall see that after
a gravitational redshift of $\Psi$, this implies that the effective
temperature $\Theta_0+\Psi = {2 \over 3}\Phi + \Psi = {1 \over 3}\Psi$
which is the famous Sachs-Wolfe (1968) result.

\titlea{3}{Primary Anisotropies and the Tight Coupling Approximation}

\def\meff{m_{\rm eff}}
Before recombination, the Compton scattering time was so short
that the photons and baryons behaved as a single fluid.  This allows
us to greatly simplify the treatment of anisotropy formation.
Specifically, since the mean free path is much shorter than
a wavelength of the fluctuation, the optical depth through a
wavelength $\sim \dot \tau/k$ is large.
Thus the evolution equations may be
expanded in $k/\dot\tau$.  Employing the baryon Euler equation
\dis\eqnBaryon,
we may eliminate the baryon velocity from the photon evolution
equation to obtain the first order equation (Peebles \& Yu 1970;
Hu \& Sugiyama 1995a)
$$
{d \over d\eta}(1+R)\dot\Theta_0 + {k^2 \over 3}\Theta_0 =
- {k^2 \over 3}(1+R)\Psi - {d \over d\eta}(1+R)\dot\Phi.
\eqn\eqnOscillator
$$
where we have dropped the higher order correction
 $\Pi_\gamma={\cal O}(k/\dot\tau)$ (see \S 3.2).
Conceptually, this equation reads: the change in momentum
of the photon-baryon fluid is determined by a competition between
the pressure restoring and the gravitational driving forces.
If we ignore the time dependence of the baryon-photon momentum
ratio $R$ from the expansion, this equation describes a forced
harmonic oscillator.  Since scattering requires the bulk
velocities of the photons and baryons to be equal, the effective
dimensionless mass of the fluid is given by $m_{\rm eff}=1+R$
to account for the inertia of the baryons.  Baryons also contribute
gravitational mass to the system as is evident in the appearance
of $m_{\rm eff}$ in the infall and dilation terms on the right hand side.
They do not however contribute significantly to the pressure or
restoring force of the system.

\titleb{3.1}{Acoustic Oscillations}

As an instructive first approximation, let us ignore time
variations in
the potentials $\Phi$ and $\Psi$ and also the baryon-photon
momentum ratio $R$ compared with changes at the oscillation frequency
$\omega=k c_s$, where the sound speed (Doroshkevich, Zel'dovich \&
Sunyaev 1978)
$$
c_s = {1 \over \sqrt{3 (1+R)}}.
$$
Equation \dis\eqnOscillator\ then reduces to the familiar form
$$
\ddot \Theta_0 + k^2 c_s^2 \Theta_0 = - {1 \over 3} k^2 \Psi.
$$
This is a simple harmonic oscillator under the constant acceleration
provided
by gravitational infall and can immediately be solved
as
$$
\Theta_0(\eta)= [\Theta_0(0)+(1+R)\Psi]\cos(kr_s)
	+ {1 \over kc_s}\dot\Theta_0(0)\sin(kr_s)
	- (1+R)\Psi,
\eqn\eqnSimple
$$
where the sound horizon $r_s = \int c_s d\eta \approx c_s\eta$.
The two initial conditions $\Theta_0(0)$ and $\dot\Theta_0(0)$ govern
the form of the acoustic oscillation.  We shall see below that they
represent the {\it adiabatic} and {\it isocurvature} modes respectively.
Equation \dis\eqnSimple\ also implies through the photon
continuity equation \dis\eqnPhoton\
that
$$
\Theta_1(\eta)=3[\Theta_0(0)+(1+R)\Psi] c_s \sin(kr_s)
		+ 3k^{-1}\dot\Theta_0(0)\cos(kr_s).
\eqn\eqnDipl
$$
In equations \dis\eqnSimple\ and \dis\eqnDipl,
lie the main acoustic and redshift
effects which dominate primary anisotropy formation.

\titlec{3.1.1}{Gravitational Infall and Redshift}

In the early universe, photons dominate the fluid and $R \rightarrow 0$.
In this limit, the oscillation takes on an even simpler form.  For
the adiabatic mode, $\dot\Theta_0(0)=0$ and $\Theta_0(\eta) =
[\Theta_0(0)+\Psi]\cos(kr_s)-\Psi$.  This represents an oscillator
with a zero point which has been displaced by gravity.  The zero
point represents the state at which gravity and pressure are
balanced.  The displacement $-\Psi > 0$ yields hotter photons
in the potential well since gravitational infall not only increases
the number density of photons but also their energy through
gravitational blueshifts.

\topinsert

\centerline{\epsfxsize=4.5in \epsfbox{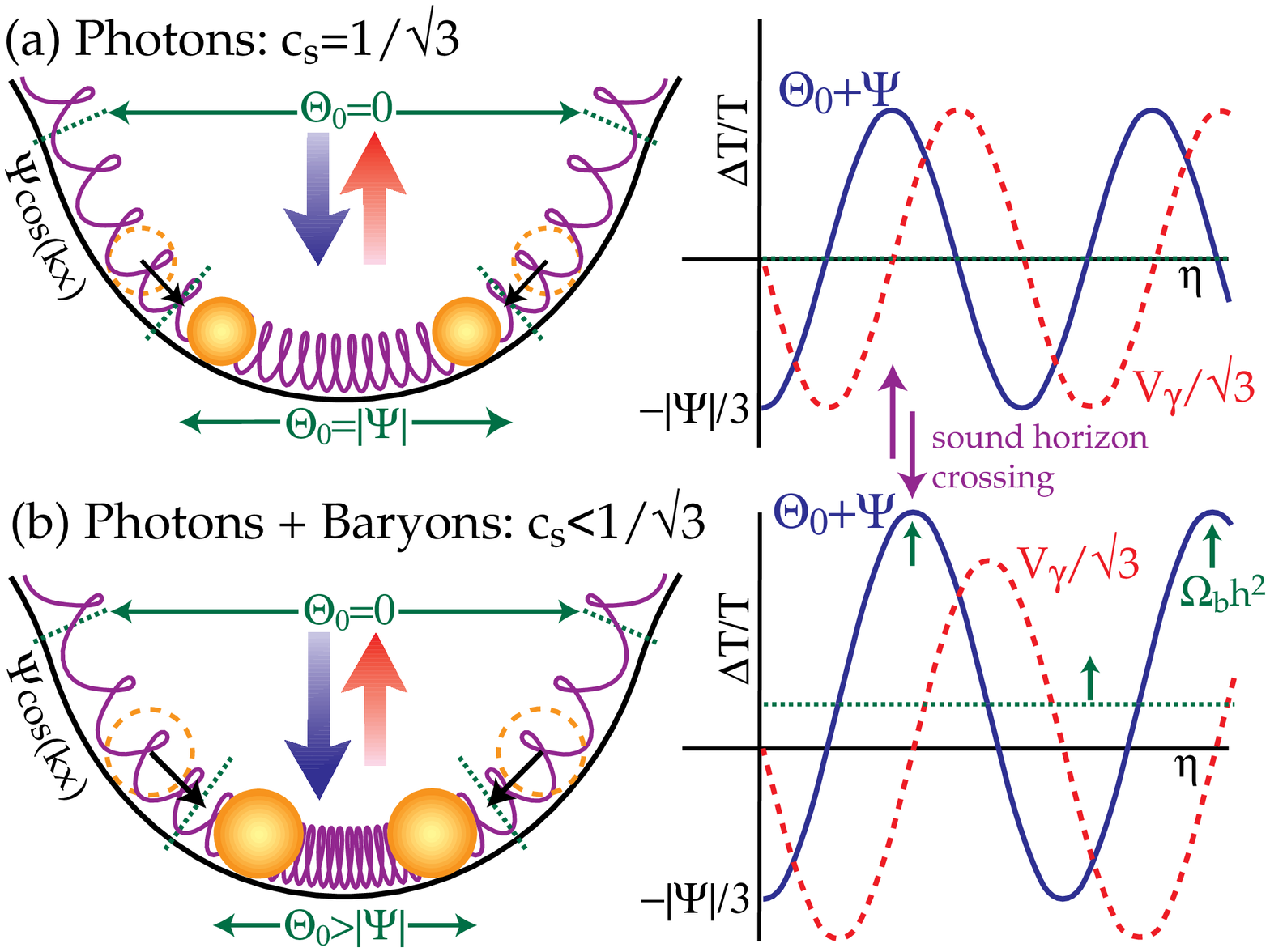}}

\vskip -1.25 truecm
\figure{2}{Acoustic oscillations. Fluid compression through
gravitational infall is resisted by photon pressure setting up
acoustic oscillations. Without baryons the photon blueshift
on infall is equal and opposite to the redshift upon leaving
the potential leading to symmetric oscillations around zero
with a Doppler effect of equal magnitude but 90 degrees out of
phase.  Baryons
increase the mass of the fluid causing more infall and a
net zero point displacement.  Temperature crests (compression)
are enhanced over troughs (rarefaction) and velocity contributions.}

\endinsert

\def\plane{e^{i {\bf k} \cdot {\bf x}}}
However, photons also suffer a gravitational redshift from climbing
out of the potential well after last scattering.  This precisely
cancels the  $-\Psi$ blueshift.  Thus, the effective temperature
perturbation is $\Theta_0(\eta)+\Psi = [\Theta_0(0)+\Psi]\cos(kr_s)$
(see Fig.~2).
The phase of the oscillation at last scattering determines the
effective fluctuation.  Since the oscillation frequency $\omega = kc_s$,
the critical wavenumber $k = \pi / r_s(\eta_*) \approx \pi /c_s\eta_*$ is
essentially at the scale of the {\it sound horizon}.  If there is
a spectrum of $k$-modes,
there will be a harmonic series
of temperature {\it fluctuation}
peaks with $k_m = m\pi/r_s(\eta_*)$ for the $m$th peak.
Odd peaks thus represent the compression phase (temperature crests),
whereas even peaks represent the rarefaction phase (temperature troughs),
inside the potential wells.

As first
calculated by Sachs \& Wolfe (1967), the effective temperature
in the matter dominated limit goes to $\Theta_0 +\Psi =
{1 \over 3}\Psi$.  In the
Newtonian frame, this is a direct consequence of the adiabatic
initial conditions.
In general, if $k\eta \ll 1$ equation \dis\eqnPhoton\ implies
$$
[\Theta_0 + \Psi](\eta_*) = [\Theta_0+\Psi](0) + [\Psi -
\Phi]\bigg|_0^{\eta_*}.
$$
In a full calculation, the small variation in the potential
at equality due to the change in the equation of state brings
the effective temperature from ${1\over 2}\Psi$ to ${1\over 3}\Psi$.
As this does not affect the qualitative picture, we will encorporate
this effect as $[\Theta_0+\Psi](0) = {1 \over 3}\Psi$ and
$\Delta(\Psi-\Phi)=0$.   Notice that if $\Theta_0(0) = \Psi(0) =
-\Phi(0) = 0$, then $[\Theta_0 + \Psi](\eta_*) = \Psi(\eta_*)-
\Phi(\eta_*) \approx 2\Psi(\eta_*)$ which yields a factor of
6 enhancement for isocurvature models.
An alternate way of deriving these results is to employ
gauge tricks as shown in \S 2.4.
The Sachs-Wolfe effect is a combination of
an intrinsic temperature and a gravitational redshift.  Since
the photon density and thus the intrinsic temperature is a gauge
dependent concept, this breakdown, but not the observable
result, will also depend on gauge.

\topinsert

\centerline{\hskip -0.75truecm \epsfxsize=3.0in \epsfbox{bdrag.epsf}}

\vskip -0.75 truecm
\figure{3}{Baryon Drag.  Baryons drag the photons into potential
wells leading to a zero point displacement of $|R\Psi|$ and
alternating peak heights in the rms fluctuation which provide
a measure of $\Omega_b h^2$.  The fall off near $a_*$ of the amplitude
is due to diffusion damping.}

\endinsert

\titlec{3.1.2}{Baryon Drag}

Though effectively pressureless, the baryons still
contribute to the inertial and gravitational
mass of the fluid $m_{\rm eff} = 1+R$.
This decreases the sound speed and changes the
balance of pressure and gravity.
Gravitational infall now leads to greater compression of the
fluid in a potential
well, \ie\ a further
displacement of the oscillation zero point (see Fig.~2).
Since the redshift is not affected
by the baryon content, this relative shift remains after
last scattering to enhance all peaks from compression over those
from rarefaction.
If the baryon-photon ratio $R$ were constant,  the effective
temperature perturbation would become
$$
\Theta(\eta)+\Psi = {1 \over 3}\Psi (1+3R)\cos(kr_s) -R\Psi,
$$
with
compressional peaks a factor of $(1+6R)$ over the $R=0$
case.
In reality, the effect is reduced since
$R\rightarrow 0$ at early times (see Fig.~3).

The {\it evolution} of the effective mass
has another effect on its own.
In classical mechanics, the ratio of energy ${1 \over 2} m_{\rm eff} \omega
^2
A^2$ to frequency $\omega$ of
an oscillator is an adiabatic
invariant.  Thus for the slow changes in $\meff \propto c_s^{-2}$,
the amplitude of the oscillation varies as $A \propto c_s^{1/2}
\propto (1+R)^{-1/4}$ since $\omega \propto c_s$.
The fundamental solutions of the oscillator
equation are modified to be
$(1+R)^{-1/4}\cos(kr_s)$ and $(1+R)^{-1/4}\sin(kr_s)$.

\titlec{3.1.3}{Doppler Effect}

Since the turning points are at
the extrema, the fluid velocity oscillates 90 degrees
out of phase with the density (see Fig.~2).
Its motion relative to the observer causes a
Doppler shift.
Whereas the observer velocity creates a pure dipole anisotropy
on the sky, the fluid velocity
causes a spatial temperature variation
$\Theta_1/\sqrt{3}$
on the last scattering surface from its line of sight component.
For a photon-dominated $c_s \approx 1/\sqrt{3}$ fluid,
equation \dis\eqnDipl\ tells us
the velocity contribution is equal in amplitude to the density
effect.
This photon-intrinsic Doppler shift should be distinguished from the
electron-induced Doppler shift of reionized scenarios.

\topinsert

\centerline{\hskip -0.75truecm \epsfxsize=3.0in \epsfbox{sep.epsf}}

\vskip -0.75 truecm
\figure{4}{Analytic decomposition ($\Omega_0=1$ scale invariant
adiabatic model).  The effective temperature after
gravitational redshift dominates the primary anisotropy.  Peak
heights are enhanced and modulated by baryon-photon ratio $R$
as well as experience a boost crossing the equality scale
at $\ell \approx \sqrt{3} (2 \Omega_0 H_0^2/a_{eq})^{1/2} \eta_0
\approx 400$.
The Doppler effect is smaller and out of phase with the
temperature.  The ISW effect due to potential decay after recombination
is small here but can be significant for low matter content universes.
Diffusion damping cuts off the acoustic spectrum at small scales.}
\endinsert

The addition of baryons significantly changes
the relative velocity contribution.  As the effective mass
increases, conservation of energy requires that
the velocity decreases for the same initial temperature displacement.
Thus the {\it relative} amplitude of the velocity scales as $c_s$.
In the toy model of a constant baryon-photon density ratio $R$,
the oscillation
becomes $\Theta_1 /\sqrt{3} = {1 \over 3}\Psi (1+3R)(1+R)^{-1/2}
\sin(kr_s)$.
Notice that velocity oscillations are symmetric around zero
unlike the temperature ones.  Thus compressional peaks will rise
clearly above
the velocity oscillations if $R$ is large.
Even in a universe with $\Omega_b h^2$ given
by nucleosynthesis, $R$ is sufficiently large
to make velocity contributions subdominant (see Fig.~4).

\titleb{3.1.4}{Acoustic Driving Effects and Isocurvature Models}

\noindent Whenever the non-relativistic matter is not the dominant
dynamical component, the potentials $\Phi$ and $\Psi$
become time-dependent. For example, when the universe is radiation
dominated,  pressure and entropy alter the behavior of the
gravitational potential.
External sources from topological
defects can also play a role.  The effects of potential evolution
can be separated into those that occur before last scattering and
thus affect the acoustic oscillations and those that occur afterwards
which affect the gravitational redshift of the photons along their
free-streaming trajectories.  We will defer consideration of
the latter to \S 4 where we consider effects between recombination
and the present.

\topinsert
\centerline{\hskip-0.5truecm \epsfxsize=2.5in \epsfbox{iso.epsf}}

\vskip -0.5truecm
\centerline{\epsfxsize=4.5in \epsfbox{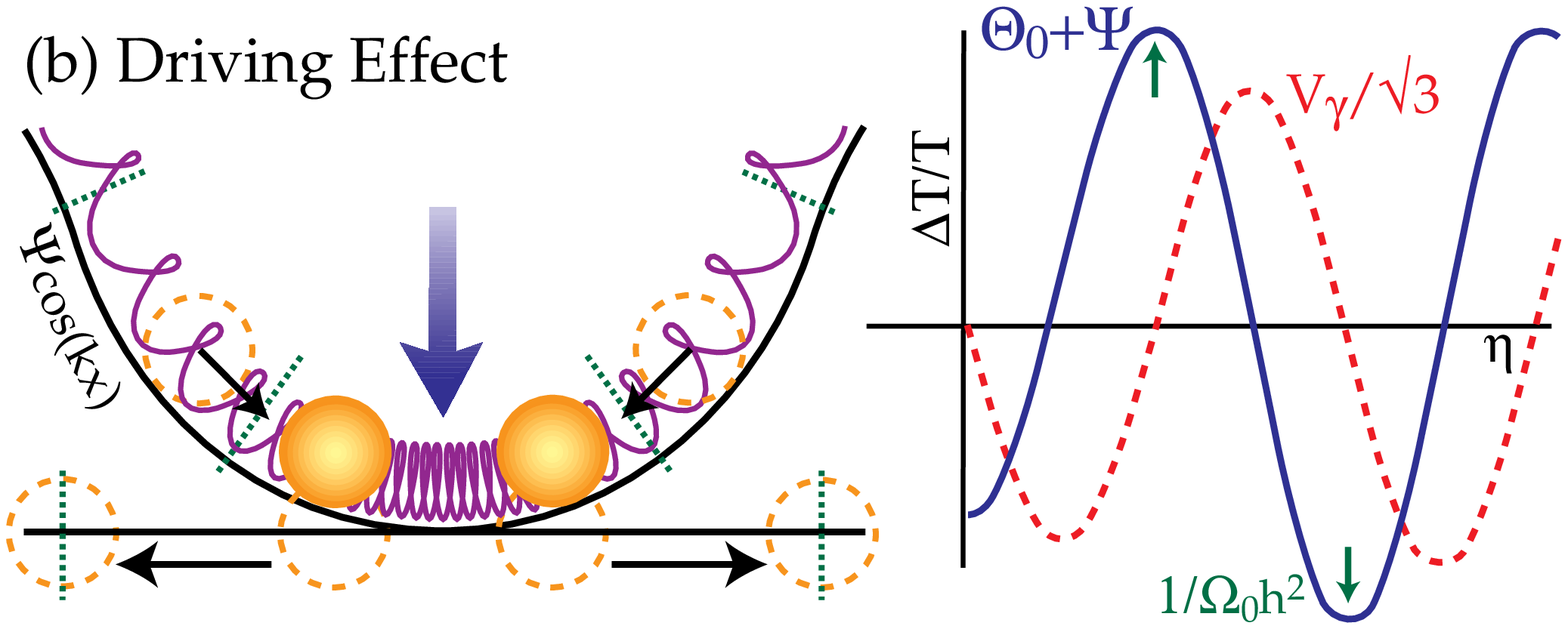}}

\vskip -0.75 truecm
\figure{5}{Driving effects.  The time evolution of the potential
can enhance the amplitude of the acoustic oscillation by
driving effects.  In the adiabatic case [(a) solid lines, numerical
results and
(b) heuristic picture], the potential
enhances the first compression through infall and decays leaving
the oscillator strongly displaced from the zero point.  In the
(baryon) isocurvature case [(a) dashed line], the potential grows from
zero and stimulates a sine mode.  The first extrema here is suppressed
due to the fact that the gravitation driving begins near
sound horizon crossing.  Notice that for the first cycle of
the adiabatic and isocurvature oscillations, the gravitational
force mimics a driving force of approximately twice the natural
period.}

\endinsert

As a simple example of the driving effects of potential evolution,
let us again consider adiabatic fluctuations where the initial
temperature $\Theta_0(0)$ and curvature fluctuations $\Phi(0)$
are finite constants related by the Poisson equation
and $\dot\Theta_0(0)=0$.
Inside the sound horizon, pressure prevents gravitational infall
in the photon-baryon and neutrino systems.
Unless CDM dominates,
energy density perturbations $\delta \rho_T$ decay with the expansion
and consequently can no longer maintain
a constant gravitational potential.
Counterintuitively, this decaying potential can actually enhance
temperature fluctuations through its near resonant driving force.
Since the potential decays after sound horizon crossing, it
drives the first compression without
a counterbalancing effect on the subsequent rarefaction stage
(see Fig.~5).

\topinsert

\centerline{\hskip -0.25truecm \epsfxsize=2.4in \epsfbox{heightsa.epsf}
 \hskip -0.25truecm \epsfxsize=2.4in \epsfbox{heightsb.epsf}}

\vskip -0.75 truecm
\figure{6}{Heights of the peaks in a scale invariant adiabatic
model.  The heights of the peaks is determined
by the baryon drag and acoustic driving effects and so are sensitive
to $\Omega_b h^2$ and $\Omega_0 h^2$. Baryons increase compression
in potential wells causing the peak heights to alternate.  Lower
matter content causes more potential decay driving the oscillation.}

\endinsert
Furthermore, the dilation effect from a decaying space
curvature $\Phi$ also enhances the acoustic fluctuations.
Heuristically, the overdensities which establish the
potential well ``stretch'' the space-time fabric.
As the potential well decays, it re-contracts.
Photons which are caught in this contraction find their wavelength
similarly contracted, \ie\ blueshifted.
Thus a differential change in $\Phi$ leads to a dilation
effect, $\dot \Theta_0 = -\dot \Phi$.  Combined, the driving effects
of infall and dilation yield an effect of order
${1 \over 3}\Psi - \Psi + \Phi \approx {5 \over 3}\Psi$ or 5
times the Sachs-Wolfe effect
when the perturbation crosses the sound horizon (see Fig.~4,5 and
Hu \& Sugiyama 1995c).
Since this effect
only occurs for modes which cross before CDM domination,
the amplitude of this boost is sensitive to the matter-radiation ratio
of the universe.  In particular, lowering $\Omega_0 h^2$, raising
the massless neutrino number or energy density, or making some of the
cold dark matter relativistic at early times through eV mass
neutrinos (see Fig.~6b;
Ma \& Bertschinger 1995; Dodelson, Gates \& Stebbins 1995),
can move the boost to larger scales and thus enhance the first few
acoustic peaks.

Similar resonant effects can occur in other situations.  For example,
many alternate scenarios follow a typical {\it isocurvature} behavior.
The hallmark of isocurvature type models is that curvature fluctuations
$\Phi$ are zero or at least suppressed outside the horizon.  The curvature
fluctuation grows by causal processes as the fluctuation crosses
the horizon.  This isocurvature behavior occurs for entropy
fluctuations where matter and radiation perturbations are
counterbalanced in the initial state, whether through
the baryons (Hu \& Sugiyama 1995b), hot dark matter (de Laix \&
Scherrer 1995) or axions (Sugiyama \& Gouda 1992),
texture models (Crittenden \& Turok 1995, Durrer, Gangui \& Sakellariadou
1995), and is at least part of the story for string models (Hindmarsh,
private communication, but see also Albrecht, Coulson \& Ferreira
1995).  The qualitative effect on isocurvature conditions is
easy to see.  Since the potential grows from zero
to a maximum near sound horizon crossing and then decays due to
radiation pressure, the force
drives the sine harmonic of oscillations (see Fig.~5a). The result
is that isocurvature acoustic peaks are 90 degrees out of
phase with their adiabatic counterparts.
The first peak occurs at larger scales that the corresponding
adiabatic peak but generally has a suppressed amplitude since
the corresponding mode is just at the sound horizon $k r_s =\pi/2$
and has not experienced the full effect of the driving term.
These features may serve to
distinguish the two basic types of scenarios in a manner
independent of the details of the given model.

Notice that in these examples, oscillations give
a coherent spectrum of peaks in integral or half integral
multiples of $\int \omega d\eta = kr_s(\eta_*)$.
Since each $k$ mode evolves independently, this coherence in
the temporal phase is the result of a special timing in the
gravitational impulse.  In all the examples considered above,
the gravitational force kicks in between horizon crossing and
sound horizon crossing.  This
explains the coherence in $k$ and the fundamental scale in
the problem $r_s(\eta_*)$.  More exotic scenarios may not
preserve this coherence.  If the source of the gravitational
potential has a more random temporal behavior, the sine
and cosine acoustic modes will be stimulated incoherently
with equal likelihood.
Combined, the two modes would lead to a smoothed out rms temperature
fluctuation below the sound horizon.  Preliminary calculations
indicate that this may be case for cosmic string models due to
complicated behavior of the string network inside the
horizon (Albrecht \etal\ 1995; Magueijo \etal\ 1995).

These considerations can be made quantitative by solving the oscillator
equation under the influence of an arbitrary but known gravitational
forcing function either analytically (Hu \& Sugiyama 1995a) or
numerically by modeling the gravitational source (e.g. Seljak 1994;
Crittenden \& Turok 1995).  Simple
closed form solutions for standard adiabatic and isocurvature models
are presented in Hu \& Sugiyama (1995c).

\titlec{3.2} {Diffusion Damping}

\noindent In reality, the photons and baryons are not perfectly coupled
since the photons
possess a mean free path in the baryons $\lambda_C \approx \dot\tau^{-1}$
due to Compton scattering.
As the photons random walk through the baryons, hot spots and cold
spots are mixed.  Fluctuations thereafter remain
only in the unscattered fraction causing a near exponential
decrease in amplitude
as the diffusion length
$\lambda_D \sim \sqrt{N}\lambda_C = \sqrt{\eta\lambda_C}$ or
$k_D \sim \sqrt{\dot\tau/\eta}$ overtakes
the wavelength.

To be more specific, diffusion causes heat conduction and generates
viscosity in the fluid (Weinberg 1972).
As photons from regions of different temperature
meet, anisotropies form leading to a quadrupole moment or
anisotropic stress in the fluid.  If the diffusion length is
well under the horizon, it overtakes the wavelength of
the fluctuation  when $\dot\tau/k = k\eta \gg 1$.  Thus the
optical depth through a wavelength is still high and the perturbative
expansion of \S 3.1 still holds.   This fact allows us to extend
the tight coupling approximation for acoustic modes to last scattering.
For small scale modes in which the optical depth through a
wavelength $\dot \tau/k$ is {\it not} high at recombination, all acoustic
oscillations will already have damped away (Hu \& Sugiyama 1995c).

Photon diffusion
is a second order effect and damps acoustic oscillations as
$\exp[-(k/k_D)^2]$ with the damping wavenumber
$$
k_D^{-2} = {1 \over 6}\int d\eta {1 \over \dot\tau}{R^2 + 4f_2^{-1}
(1+R)/5 \over (1+R)^2},
$$
where $f_2$ accounts for the subtle effects in
the generation of anisotropic stress (see Fig.~4).
Kaiser (1983) showed that
 $f_2 =9/10$ due to the angular dependence and $3/4$ if the additional
effects of polarization are included.  The two processes aid the
generation of the quadrupole moment and hence increase the viscous
damping of the acoustic oscillations.

The baryons are dynamically coupled to the photons by momentum
exchange in Compton scattering from $ \dot\tau_d \equiv
\dot \tau/R$ in equation \dis\eqnBaryon.
If $\dot \tau_d/k \gg 1$,
the baryons will be dragged in and out of potential
wells with the photons.  This process destroys the baryonic
acoustic oscillations as well and is known as Silk (1968) damping.
Since coupling requires coevolution in the number density
$\dot \delta_b = {3 \over 4} \dot \delta_\gamma
= 3\dot\Theta_0$, only entropy fluctuations $S = \delta_b - {3 \over 4}
\delta_\gamma$ survive diffusion damping.

Notice also that the ionization history factors in the diffusion length
through the mean free path $\dot\tau^{-1}$.
As we shall now see, at recombination the mean free path
and hence the diffusion length increases substantially but does
not approach the horizon scale.

\titlec{3.3} {Decoupling}

The CMB anisotropy today is
simply the acoustic fluctuation at last scattering, modified by diffusion
damping and free streamed to the present:
$$
[\Theta+\Psi](\eta_0,k,\mu) \approx [\Theta_0 + \Psi - i\mu \Theta_1]
(\eta_*,k)
{\cal D}_\gamma(k) e^{ik\mu(\eta_*-\eta_0)},
\eqn\eqnFS
$$
for flat space, where ${\bf k} \cdot \bg = k\mu$,
$$
{\cal D}_\gamma(k)=\int_0^{\eta_0} d\eta \, {\cal V}_\gamma
e^{-[k/k_D(\eta)]^2},
\eqn\eqnDamp
$$
is the diffusion damping factor
and the visibility function ${\cal V}_\gamma = \dot \tau e^{-\tau}$
is the probability of last scattering within $d\eta$ of $\eta$.
Its peak is near $\tau(z_*) = 1$.
The damping factor is dependent only on the background cosmology
and fits across the whole range of parameter space are presented
in Hu \& Sugiyama (1995c).  We shall examine the meaning and
implications of each of these terms below.

The baryons do not decouple from the photons precisely at last
scattering.  As we have seen in \S 3.2, the coupling strength
is altered by a factor of $R$, $\dot\tau_d = \dot\tau/R$.  By
analogy to the photon case, we can define a drag optical
depth $\tau_d$ and the end of the drag epoch as $\tau_d(z_d) = 1$
(Hu \& Sugiyama 1995c).
After this point, Compton drag on the baryons can no longer prevent
the gravitational infall of the baryons.  Because $R(z_*) \sim
0.3$ for standard recombination and a big bang nucleosynthesis
value for $\Omega_b h^2$, $z_*$ and $z_d$ approximately coincide
for the standard case.   However for reionized scenarios
$R(z_*) \gg 1$ and the baryons decouple from the photons long
before last scattering.  It is thus no longer appropriate
to consider the photons and baryons as tightly coupled at last
scattering.   We will develop new techniques to handle this
situation in \S 4.   From the drag depth $\tau_d$, a drag
visibility function ${\cal V}_b \approx \dot\tau_d e^{-\tau_d}$
can be constructed
(see Hu \& Sugiyama 1995c for the small correction due to
expansion damping).
The final scale for Silk damping is obtained from the visibility
function and the diffusion length as in equation \dis\eqnDamp\ for
the photons.

\topinsert

\centerline{\epsfxsize=4.0in \epsfbox{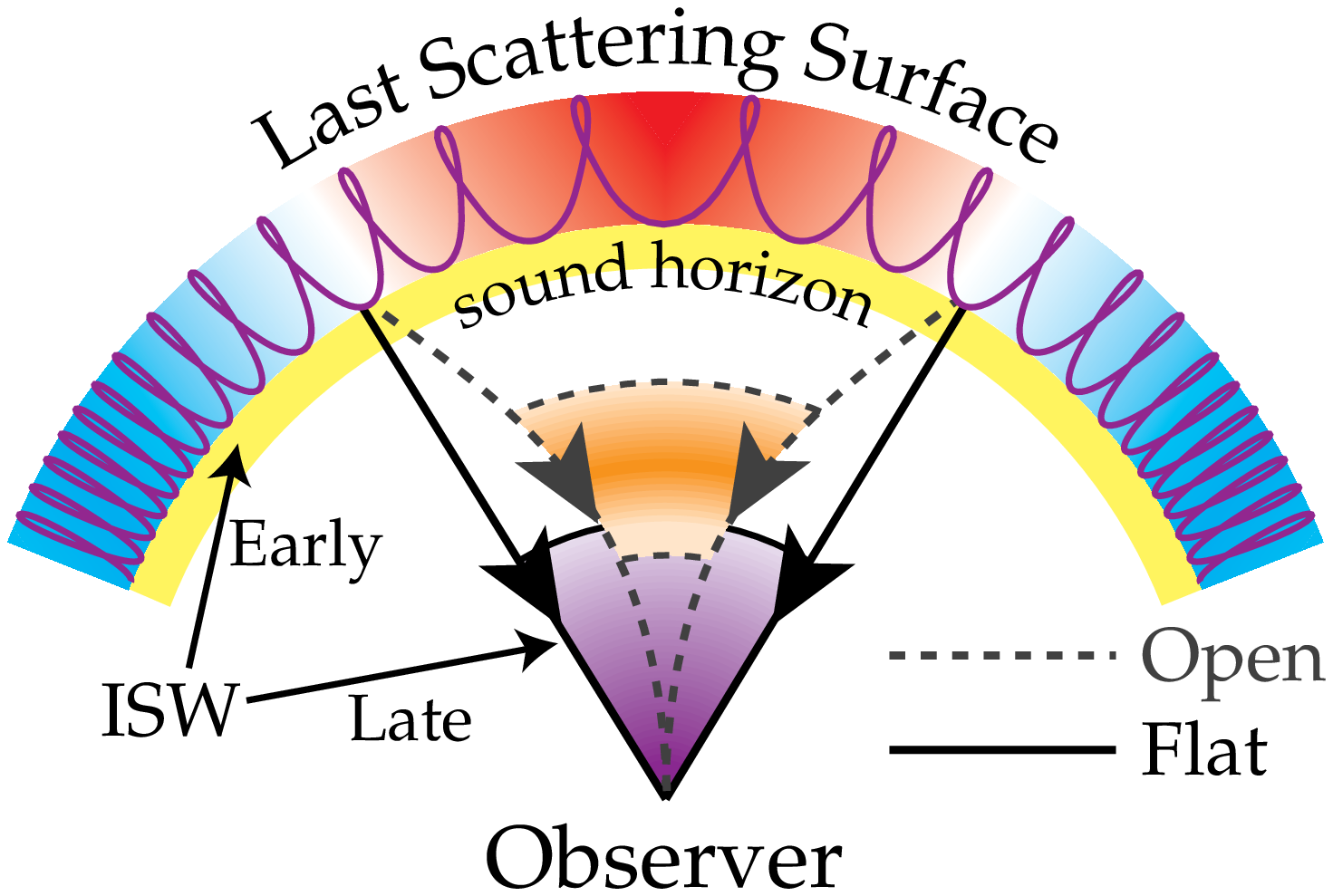}}

\vskip -0.5truecm
\centerline{\hskip -0.25truecm \epsfxsize=2.4in \epsfbox{projc.epsf}
 \hskip -0.25truecm \epsfxsize=2.4in \epsfbox{projb.epsf}}

\vskip -0.75 truecm
\figure{7}{Location of the peaks. The location of the peaks is
determined by the angle the sound horizon subtends at last scattering.
In an open universe, geodesic deviation makes this angle much smaller
than in a flat universe.  As this
angle involves a ratio of the sound horizon to the (angular
diameter) distance to
last scattering, distance scale changes through $\Lambda$ and $h$
have little effect on its angle.
The drop in the lowest multipoles for the open
$\Omega_0=0.1$ is due to the lack
of supercurvature scale power in the otherwise
scale invariant initial conditions assumed here. Models here
have $h=0.5$ and $\Omega_b=0.05$.
}

\endinsert

\titlec{3.4} {Projection Effects and the Anisotropy Spectrum}

The presence of the $\exp(ik\mu\Delta\eta)$ term in equation
\dis\eqnFS\ represents the free streaming of the photons in
flat space.
Photons travel at the speed of light so that the number of
wavelengths traveled between last scattering and today
is $k\Delta\eta$.  However, the phase of the wavefront changes
only in the perpendicular direction so that the observed
phase change will depend on the line of sight
$\phi = k\Delta\eta\mu$.
Since inhomogeneities
at the last scattering surface appear as anisotropies on the sky
today, this can equivalently be viewed as a simple projection
of the plane wave on the sphere.
The anisotropy is expressed through the decomposition of the
plane wave into multipole moments
$\ell \sim \theta^{-1}$,
$$
{\Theta_\ell(\eta_0,k) \over 2\ell+1} = \left\{
[\Theta_0 + \Psi](\eta_*,k) + \Theta_1(\eta_*,k){1 \over k}
{d \over d\eta} \right\} {\cal D_\gamma}(k) j_\ell(k\Delta\eta) .
\eqn\eqnThetaL
$$
Since $j_\ell$ peaks at $\ell \sim k\Delta\eta$, it is clear that
free streaming just projects the physical scale $k \sim \lambda^{-1}$
onto an angular scale as $\theta \sim \lambda/\Delta\eta$.
The total power in the $\ell$th multipole is obtained by
integration over $k$ modes, $C_\ell = (2/\pi)\int
k^3 |\Theta_\ell(\eta_0,k)|^2/(2\ell+1)^2 d\ln k$ .

\topinsert

\centerline{\hskip -0.25truecm \epsfxsize=4.5in \epsfbox{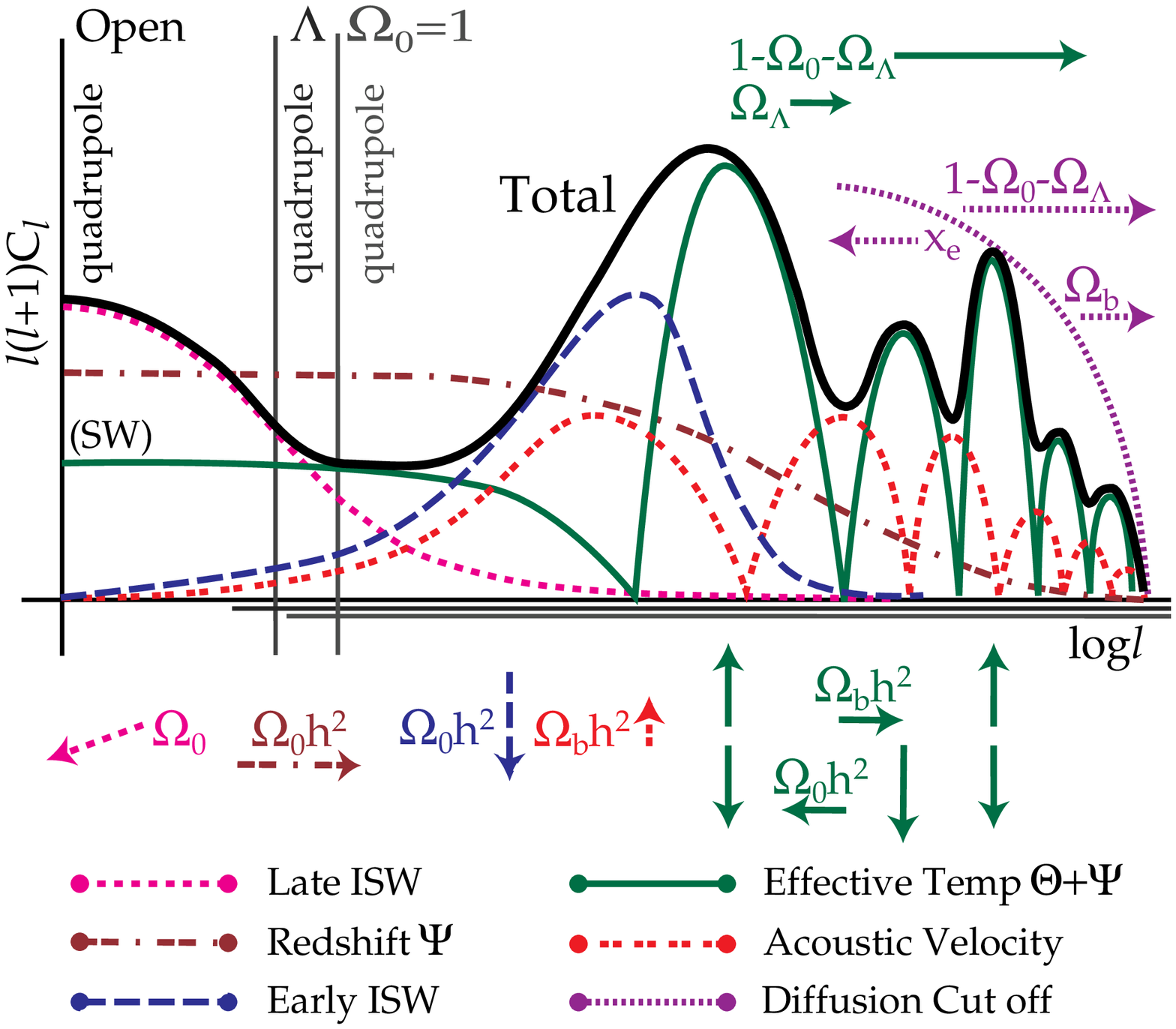}}

\vskip -0.75 truecm
\figure{8}{Parameter sensitivity.  A schematic representation based
on a scale invariant adiabatic scalar model.  Features in
an open model are shifted to significantly smaller angles compared with
$\Lambda$ and $\Omega_0=1$ models, represented here as a shift
in the $\ell$ axis. Isocurvature models behave similarly with
respect to the acoustic oscillations save that the peaks are
90 degrees out of phase.  The spectrum may also be tilted by changing
the initial power spectrum from the scale invariant assumption.}

\endinsert

The physical content of the generalization to open universes is
now obvious.  Since free streaming merely represents a projection,
one replaces the comoving distance $\Delta\eta$ with
the comoving angular diameter distance $r_\theta
= |K|^{-1/2} \sinh[|K|^{1/2}\Delta\eta]$, where
the curvature $K = -H_0^2(1-\Omega_0-\Omega_\Lambda)$.   For
a $K > 0$ closed geometry, merely replace the sinh with sin.
Notice that in the closed case, the location of the first peak
has oscillates with the curvature (White \& Scott 1995).
Putting this
together, the acoustic peaks occur at
$$
\ell_j = k_j |r_\theta(\eta_*)| = \Big| {r_\theta \over
	r_s } \Big|_{\eta_*} \times \cases{ j\pi, & adiabatic \cr
				   (j-1/2)\pi. & isocurvature \cr }
$$
Formally, this arises by replacing $j_\ell$ with the radial eigenfunction
of an open universe (Wilson 1983).
Notice that the peak locations
only depend on the background
cosmology once the adiabatic or isocurvature nature of the fluctuations
has been established.  Indeed, the ratio of the peak locations themselves
can be employed to separate the two (Hu \& Sugiyama 1995b).
The dominant factor in the peak locations is the curvature of
the universe which can make the same physical scale subtend a much
smaller angle on the sky (see Fig.~7).
Here we have a classical cosmological test of the
curvature: knowing the physical scale (sound horizon at last scattering)
and the redshift ($z_* \approx 1000$) of the
acoustic peak, we measure its angular extent.  The inferred angular
diameter distance tells us the curvature.  If the peak location
is known to sufficient accuracy,  we can also measure $\Omega_b h^2$,
$\Omega_0 h^2$ and $\Lambda$ through their effects on the sound
horizon and angular diameter distance (see Figs. 6 \& 7).
Analogous information can be obtained from the location of the
diffusion damping scale even if oscillations are not apparent due
to random driving effects.

The heights of the peaks also contain important, if less model independent
information. As we have seen,
every other peak is boosted by the baryon content due to baryon
drag and scales inside the horizon at equality probe $\Omega_0 h^2$
or more generally the matter-radiation ratio which includes
information on the neutrino content and mass.  Since these two
effects have relatively robust features, alternating peak heights
and a strong height boost at matter-radiation equality, there is
hope of obtaining information out of the heights
of the peaks as well as their location for a wide class of
models.  As an example, we present
a schematic picture for the anisotropy spectrum in Fig. 8 based on
the popular scale invariant adiabatic model.
Note that changing the overall dynamics from $\Omega_0=1$ through
flat $\Lambda$ models to open models is similar to shifting the spectrum
in angular space toward smaller angles.

\titlea{4}{Secondary Anisotropies and the Weak Coupling Approximation}

We have been assuming up to this point that CMB fluctuations are
frozen in at a redshift $z_* \approx 1000$.  Several processes
in the foreground of recombination could alter the anisotropy
spectrum.  These are generally known as {\it secondary} anisotropies.
Just as the tight coupling approximation assisted in the calculation
and interpretation of primary anisotropies, the weak coupling
approximation helps in understanding secondary anisotropies
(Hu \& White 1995).

In abstract form, anisotropy generation is governed by internal
sources from photon fluctuations at last scattering $S_{prim}$
and sources external to the photon system in the foreground
of recombination $S_{sec}$ by
$$
{\Theta_\ell(\eta_0,k) \over 2\ell+1} =
\int_0^{\eta_0} [S_{prim} + S_{sec}] j_\ell(k\Delta\eta)d\eta,
\eqn\eqnAbstractProjection
$$
in flat space with an appropriate generalization of the radial
eigenfunction for an open geometry (Wilson 1983). Equation
\dis\eqnAbstractProjection\ just states that we observe the projection
of the source at $\eta$ on a shell at a distance $\Delta\eta
=\eta_0-\eta$.
Whether the sources or the phase of the wave varies more
rapidly distinguishes the tight from the weak coupling
regime.  For primary anisotropies, the source is localized
over a short range of time around recombination.
Since $\dot\tau/k \gg 1$, $j_\ell$ can be taken out of the integral
\dis\eqnAbstractProjection\ leading to the simple projection of
equation \dis\eqnThetaL.

Secondary anisotropies
do not necessarily possess this property.   Most cosmological
effects aside from decoupling take on the order of an expansion
time at the relevant epoch to be completed. In these cases,
$S_{sec}$ may be taken to be slowly varying at small scales
and removed from the integral in
\dis\eqnAbstractProjection.   Since
$$
\int_0^{\eta_0}
j_\ell(k\Delta\eta) d\eta  \approx {\sqrt{\pi} \over 2k}
{\Gamma[{1 \over2}(\ell + 1)] \over \Gamma[{1 \over 2}(\ell+2)]} \approx
\sqrt{\pi \over 2\ell}{1 \over k},
\eqn\eqnWeak
$$
in the weak coupling limit where $\dot S_{sec}/(k S_{sec}) \ll 1$,
anisotropies fall with $\ell$ more rapidly
than a simple projection of the source would imply.
This just reflects the fact that it contributes
across many wavelengths of the fluctuation allowing contributions
from crests and troughs to cancel.   We will now discuss the
main sources of secondary anisotropies $S_{sec}$.

\titleb{4.1}{Gravitational Effects}

Even in the absence of reionization, secondary anisotropies can be
generated by gravitational redshift effects between recombination
and today.
The differential redshift from $\dot \Psi$ and dilation from $\dot \Phi$
discussed above must be integrated along the trajectory
of the photons $S_{sec} = \dot\Psi - \dot \Phi$.   Tensor
fluctuations can also give rise to anisotropies through this
mechanism.
We thus call the combination the {\it integrated}
Sachs-Wolfe (ISW) effect.  Notice that these effects only
occur if the metric fluctuation is time varying.  We can
separate this general
mechanism for anisotropy formation into categories based on the
reason for the time evolution.  There are four possibilities
to consider: time evolution due to the radiation content (early ISW
effect), due to the expansion (late ISW effect), due to non-linear
evolution (Rees-Sciama effect) and any more exotic sources such
as gravity waves or defects (sourced ISW).
Additionally, beyond linear theory gravitational lensing may affect the
CMB.  Since this merely shuffles power in anisotropies around scales
at the arcminute level for a CDM type model, we will not further
consider the effect
and refer the interested reader to Seljak (1995a) and references
therein.

\titlec{4.1.1} {Early ISW Effect}

The early ISW effect is the direct analogue of the acoustic driving
effect of \S 3.1.4 except that the photons are in the
free streaming rather than the
tight coupling regime.  For adiabatic conditions, the potential
decays after horizon crossing in the radiation dominated limit.
For isocurvature conditions, it grows outside the horizon and
then decays as in the adiabatic case.  In general, this effect
will smoothly match onto the acoustic peaks due to similarities
in the cause of their generation.  Due to the
later time of generation, this effect influences larger
scales than the acoustic peak but is cut off above the equality
scale.  Furthermore, since it arises from a distance closer
to ourselves, the same physical scale subtends a larger angle
on the sky.  Together these considerations imply that the
early ISW effect fills in the anisotropy on scales just larger
than the first acoustic peak (see Fig.~8).  It serves to broaden the rise
and shifts the location of the first peak in the spectrum
to larger scales.  Unfortunately, this effect satisfies
neither the tight nor the weak coupling approximation
since the decay time $\sim \eta$ and the wavelength $ \sim k^{-1}$
are by definition comparable for this horizon crossing effect.

\titlec{4.1.2} {Late ISW Effect}

If the universe has a non-vanishing curvature or cosmological constant,
eventually these will dominate the expansion rate.  Under the rapid
expansion, density fluctuation $\delta\rho_T$ decays taking the
potential with it.
The decay takes on the order of an expansion time at the end of
matter domination independent of the wavelength.
Since the photons
free
stream, they travel across many wavelengths of
the perturbation on scales smaller than the horizon.
If the potential decays while the photon
is in an underdense region, it
will suffer an effective redshift rather than a blueshift.
Contributions from overdense and underdense regions will
cancel and damp the ISW effect on small scales.
This is the hallmark of the weak coupling regime and is mathematically
expressed through cancellation in the integral \dis\eqnWeak.

For a fixed $\Omega_0$, the decay epoch occurs much later in
flat $\Omega_\Lambda + \Omega_0 = 1$ models than open ones.
Thus $\Lambda$ models will suffer
cancellation of late ISW contributions at a much larger scale than open
models.  In fact, for reasonable $\Omega_0 \simgt 0.1$ decay the
decay has already started at the quadrupole (see Fig.~7,8;
Kofman \& Starobinskii
1985; Hu \& White 1995).  In summary,
the epoch that the universe exits the matter
dominated phase is imprinted on the CMB by the late ISW effect.

\titlec{4.1.3} {Rees-Sciama Effect}

Even for an $\Omega_0 = 1$ CDM dominated universe, potentials only remain
constant in linear perturbation theory.  The second order contribution
has been shown to be negligibly small (Martinez-Gonzalez, Sanz \& Silk
1992).  The effect in the non-linear regime  has been estimated with
N-body simulations through power spectrum techniques (Seljak 1995)
and ray tracing techniques (Tuluie, Laguna \& Anninos  1995).
The general conclusion is that this effect does not become
comparable to the primary signal until well into the diffusion
damping tail. In the absence of reionization, it should
therefore present no problem for the extraction of main features
such as the curvature and the baryon content from the acoustic
peaks.   It may however complicate the extraction of cosmological
parameters from more subtle effects, for example the neutrino mass.

\titlec{4.1.4} {Sourced ISW Effect}

This catch-all category contains all other gravitational redshift
effects on the CMB.
Tensor perturbations in the metric, \ie\ gravity waves,
give rise to dilation effects as the photons free stream.
Due to the nature of the metric distortion, they leave a quadrupole
signature in the CMB which thereafter is projected onto
higher multipoles.  Combined with cancellation effects, this
implies that the tensor spectrum thus typically exhibits
a sharp drop in power from the quadrupole.  Since contributions
only arise between last scattering and the present, there is
also a cut off at the angle the horizon subtends at last scattering.
Below this scale, the gravity waves have already redshifted
away by last scattering.  Detailed calculations of the
spectrum were first carried out by Crittenden \etal\ (1994)
and the spectrum shown in Fig.~9 is from Hu \etal\ (1995).
For inflationary models, there exists a consistency relationship
between the amplitudes of the scalar and tensor modes and the
slope of the power spectrum (see \eg\ Steinhardt 1995).  There
is hope that detailed measurements of the CMB spectrum can
thus provide a strong test of the inflationary scenario
(see Lindsey \etal\ 1995 and references therein).

\topinsert
\centerline{\hskip -0.75truecm \epsfxsize=3.0in \epsfbox{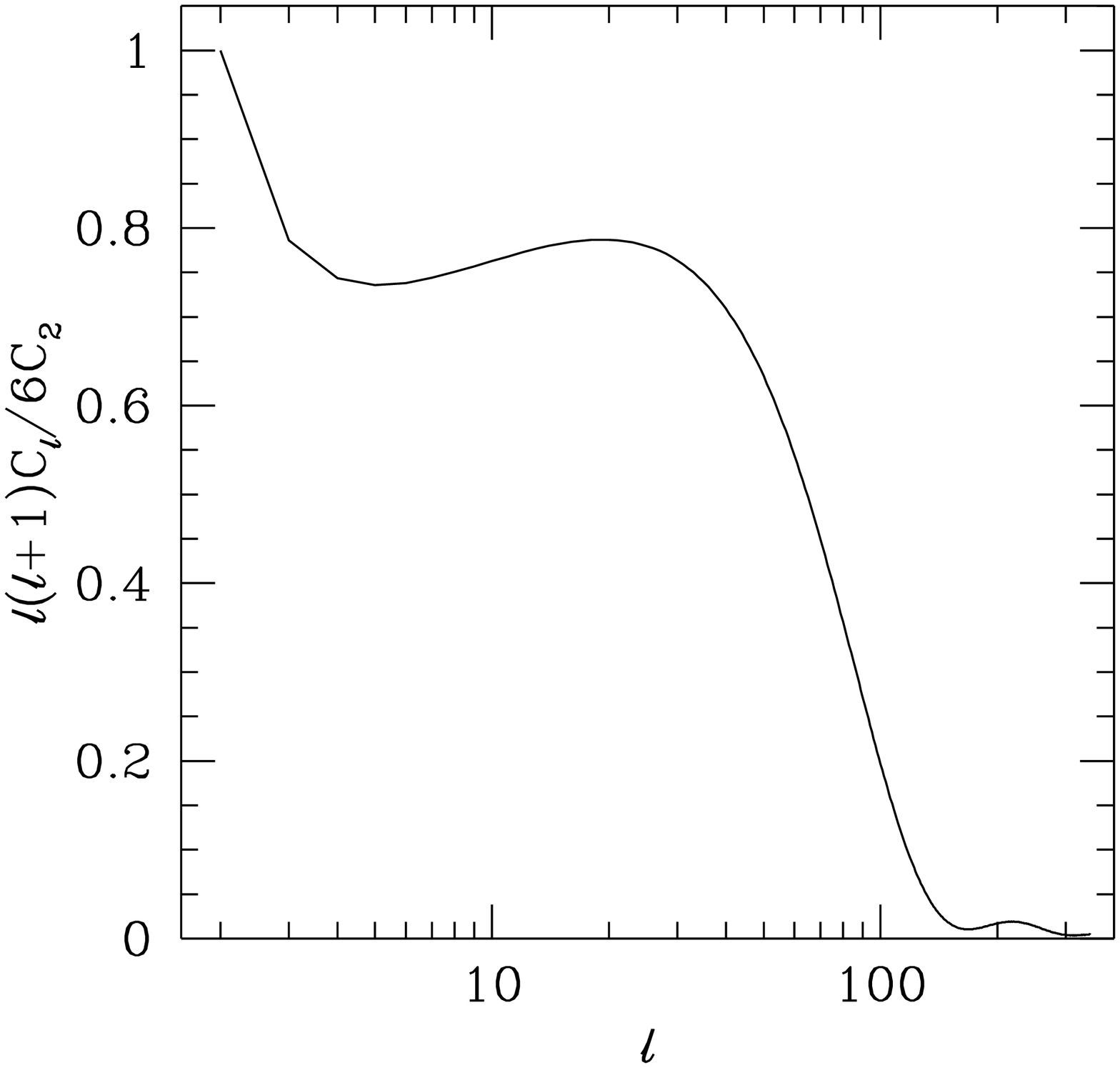}}
\vskip -0.75 truecm
\figure{9}{Gravitational Wave ISW Effect.  The metric distortion
caused by a gravitational wave induces a redshift/dilation effect
on CMB photons.  The spectrum is cut off below the angle subtended
by the horizon at last scattering since gravity waves only affect
free streaming photons significantly.  Due to the signature of the
metric distortion, it originates as a source of the quadrupole anisotropy.}
\endinsert

Cosmological defects may also act as an external
source of gravitational potential perturbations.  Pen, Spergel \&
Turok (1993) find that in such models a scale invariant form for
the anisotropy results from this effect and can mimic the inflationary
prediction.   As in the case of the late ISW effect, the magnitude
drops from $2\Psi$ to zero as the characteristic time scale for the
change in $\Psi$ becomes longer than the light travel time
across a wavelength.  For defect models Pen, Spergel  \& Turok (1993)
estimate the amplitude as approximately ${4 \over 3} \Psi$.
Since this effect occurs around horizon crossing for a given mode
due to the isocurvature conditions, the largest spatial modes have
not been enhanced and can lead to a drop in power at the lowest
multipoles.

\titleb{4.2}{Scattering Effects}

An early round of structure formation may be able to reionize all
or part of the universe at high redshift.  If this occurs
at $$z \simgt z_* \approx  10^2 (\Omega_0 h^2 /0.25)^{1/3}
(x_e \Omega_b h^2 /0.0125)^{-2/3},$$ where $\tau(z_*)=1$
 then CMB anisotropies will be
drastically influenced by scattering.  The model-independent
information stored in the primary anisotropy spectrum will be
greatly reduced. It is replaced by more detailed information
and clues about the evolution of structure in the universe.

\topinsert
\centerline{\hskip -0.75truecm \epsfxsize=3.0in \epsfbox{vishn.epsf}}
\vskip -0.75 truecm
\figure{10}{Reionization damping, Doppler, and Vishniac effects
(scale invariant adiabatic $\Omega_0=1$, $\Omega_b=0.05$, $h=0.5$,
COBE normalized model).
As the ionization level between recombination and the present increases,
primary anisotropies are reduced by diffusion and rescattering.
For sufficiently high ionization, here $z_i \simgt 100$, the Doppler
effect on the new last scattering surface can regenerate some
fluctuations at intermediate scales.  The Vishniac effect dominates
at very small angular scales.  Due to its second order nature, it
contributes even if the ionization redshift is low. The
$z_i=5,10$ curves are indistinguishable for the primary anisotropies
as are the $z_i=100,1000$ for the Vishniac effect.}
\endinsert

\titlec{4.2.1}{Reionization Damping}

The most dramatic effect of reionization is the increase
in the photon diffusion length.  Since last scattering is delayed
until the Compton mean free path approaches the horizon
(or equivalently when the scattering rate $\dot \tau$
drops below the expansion rate $\dot a/a$), the diffusion
length at last scattering is the horizon scale $\lambda_D
\approx \sqrt{\eta/\dot\tau} \approx \sqrt{\eta a/\dot a} \approx \eta$.
Intrinsic photon
fluctuations such as the acoustic oscillations will be damped by
diffusion below the horizon scale.  Moreover, since acoustic
oscillations appear only below the sound horizon no oscillations will
be apparent in the reionized spectrum.

If reionization is not sufficiently early, some trace of the acoustic
oscillations may remain in the spectrum.  Recall that diffusion damping
works since rescattering of photons arriving from directions with
different  intrinsic temperatures varies destroys the anisotropy.
Fluctuations
are only retained in the unscattered fraction $e^{-\tau}$.  If
the total optical depth between recombination and the present is
$\tau \simlt 1$, primary and secondary scattering anisotropies
may be present in the spectrum (see Fig.~10).
In this case, the information contained
in the CMB would truly be enormous but also difficult to extract.

\titlec{4.2.2}{Cancelled Doppler Effect}

After the Compton drag epoch ${3 \over 5} z_d \approx 165
(\Omega_0 h^2)^{1/5} x_e^{-2/5}$ where $\tau_d(z_d)=1$
(Hu \& Sugiyama 1995c), baryonic
gravitational instability can no longer be prevented by the photons.
Collapse of the baryon density fluctuations implies that baryon
peculiar velocities will create a Doppler effect in the CMB.
However the weak coupling approximation tells us that these too
will be damped as the photon traverses many crests and troughs
of the perturbation at last scattering.  In fact, cancellation
for the Doppler effect is particularly severe.  If the perturbation
wavevector is perpendicular to the line of sight ${\bf k} \perp
\bg$,  cancellation
is avoided.  However, flows are irrotational in linear theory
so ${\bf v}_b \parallel {\bf k}$.  Since
the velocity is then perpendicular to the line of sight, no
Doppler effect arises $\bg \cdot {\bf v}_b = v_b \bg \cdot
\hat {\bf k} = 0$ (see Fig.~11).
Fluctuations only survive if there are
variations in the velocity or optical depth through last scattering.
Thus the Doppler source is related to the derivative of these
quantities as
$
S_{sec} = e^{-\tau}[\dot V_b \dot\tau + V_b \ddot \tau]/k
$
at small scales where
recall ${\bf v}_b({\bf x}) = -i V_b \plane \hat {\bf k}$.
This leads to a suppression by a factor of approximately
$(k\eta_*)^{-1}$ (Kaiser 1984).
Cancellation under the weak coupling approximation
yields a further suppression via equation \dis\eqnWeak.  In general,
later last scattering implies a greater $k\eta_*$, a larger
cancellation scale, and hence smaller Doppler fluctuations.
Nevertheless, in Fig.~10 the slight upturn at intermediate
scales for the high
ionization case is due to this effect.

\topinsert
\vskip -0.5 truecm
\centerline{\hskip -3.5 truecm\epsfxsize=3.5in \epsfbox{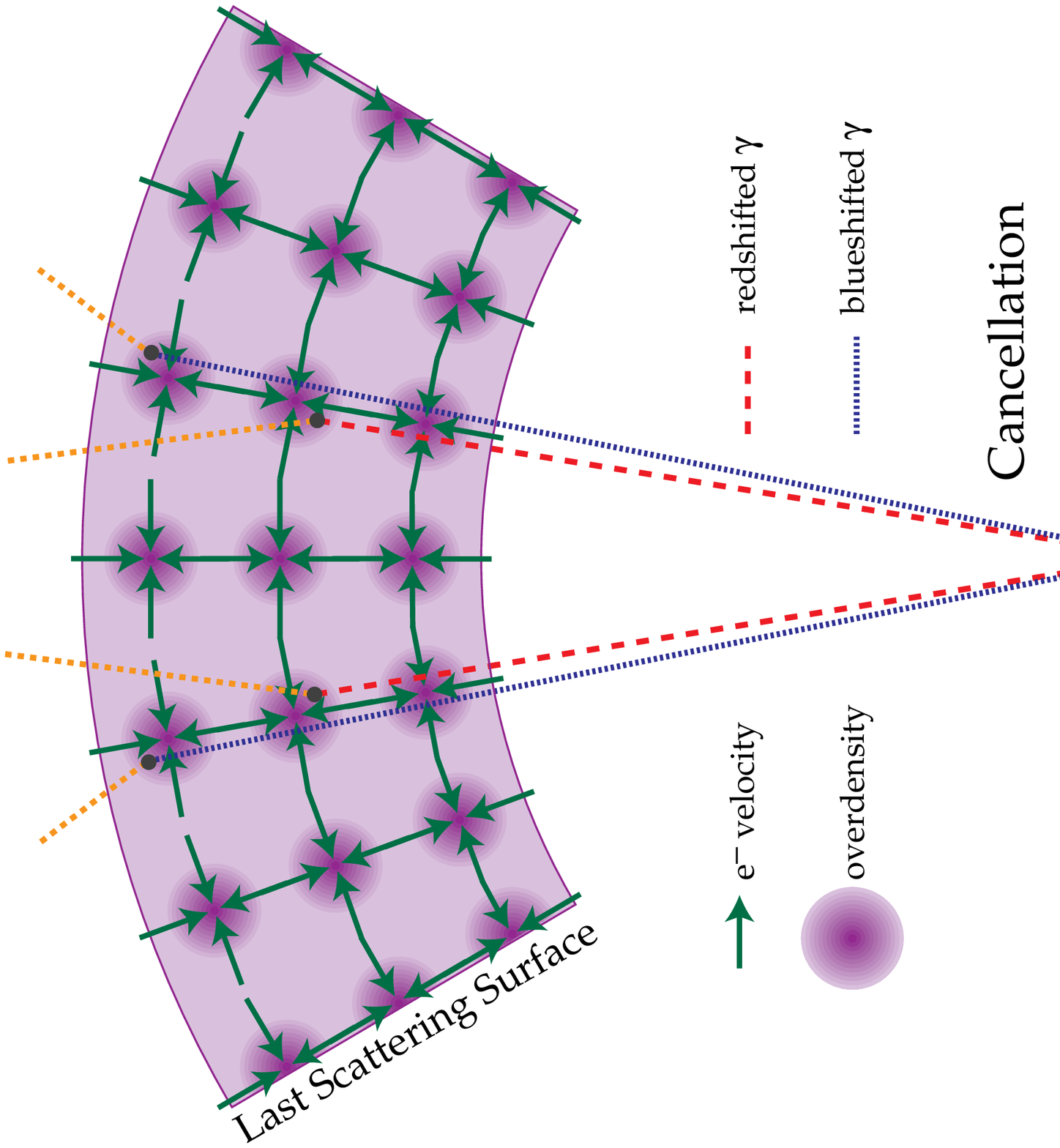}}
\vskip -1.2 truecm
\centerline{\hskip -3.85 truecm\epsfxsize=3.5in \epsfbox{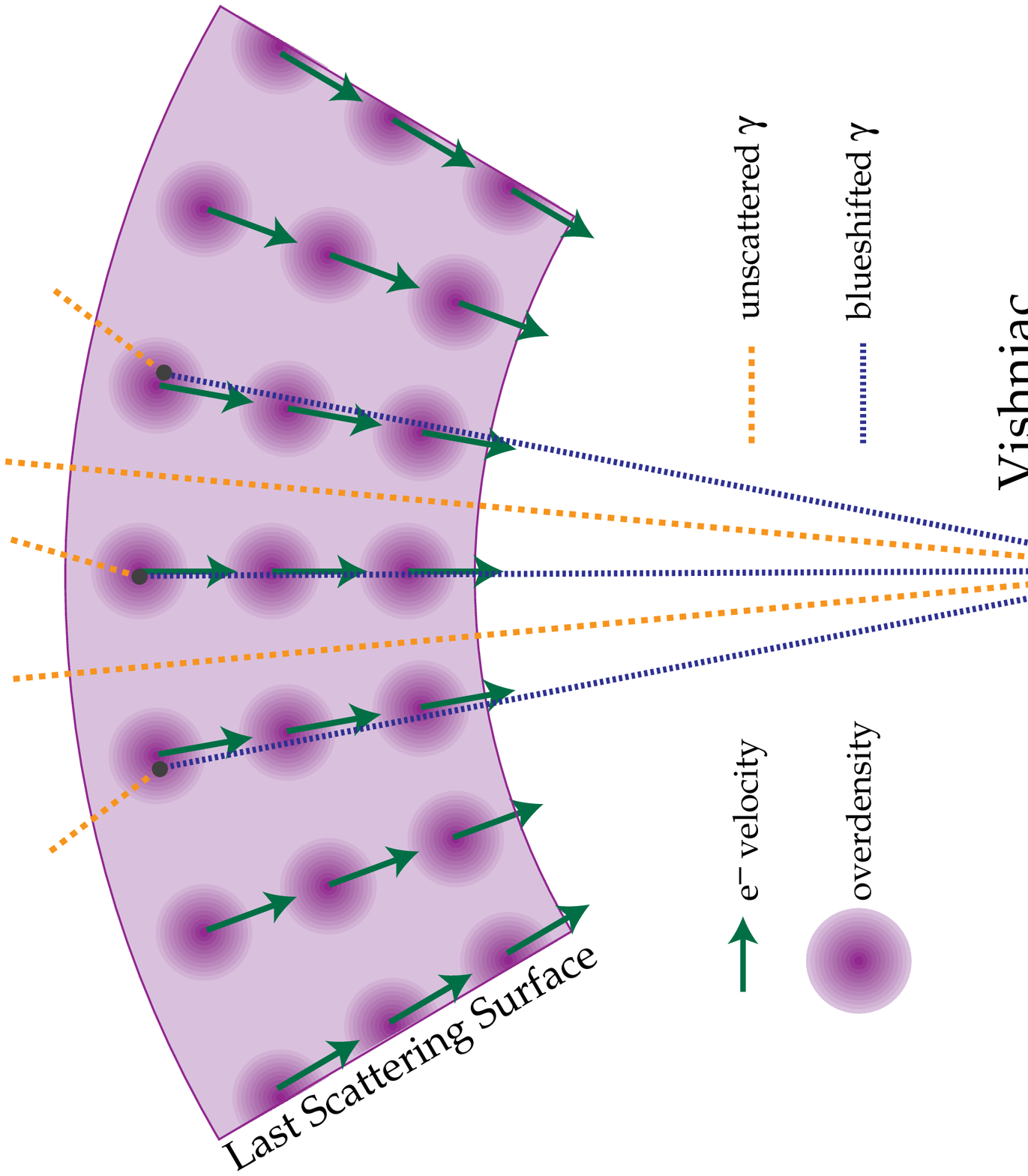}}
\vskip -1.5 truecm
\figure{11}{Cancellation vs. Vishniac mechanisms}

\endinsert

\titleb{4.2.3}{Vishniac Effect}

Since the first order Doppler fluctuation is severely suppressed
due to ``coincidental'' geometrical reasons other effects may
dominate the anisotropy at small scales.
Variations in the optical depth due to density fluctuations
across the last scattering surface
may alow the Doppler effect to escape severe cancellation.
The increased probability of scattering
causes a preferential generation of Doppler fluctuations in
overdense regions.  The ensuing anisotropy is not as severely damped
as the first order contribution.  This is because the velocity field
can be parallel to the line of sight while the density variation is
perpendicular to the line of sight (see Fig.~11).
This effect can also be calculated
under the weak coupling approximation.  It is slightly more complicated
than linear effects since it entails mode coupling between the density
and velocity fields.  An explicit
expression is given in Hu \& White (1995).  For a CDM model, we
find that Vishniac contributions do indeed exceed linear contributions
at sufficiently small scales.  Furthermore, since second order effects
are strongly peaked to late times, even minimally ionized scenarios
carry a significant fraction of the maximum signal (see Fig. 10).

\titleb{4.2.4}{Cluster Sunyaev-Zeldovich Effect}

Clusters provide a non-linear analogue of the Vishniac effect.
Here the hot cluster provides the variation in the optical depth
which causes preferential scattering.  The Doppler effect due
to the peculiar velocity of the cluster yields an anisotropy
known as the {\it kinematic} Sunyaev-Zel'dovich (SZ) effect
(Sunyaev \& Zel'dovich 1972).
For an individual cluster the temperature fluctuation is of
order $\tau_c v_c$, where the optical depth through a cluster
is of order $\tau_c \approx 0.1 - 0.01$ and the peculiar velocity
$v_c \approx {\rm few} \times 10^{-3}$.  This provides
an interesting way of measuring the peculiar velocity of
a cluster without introducing the problems associated with
determining the distance scale (see \eg\ Sunyaev \& Zel'dovich 1980,
Haehnelt \& Tegmark 1995).  The average effect is
much smaller however and probably does never dominates the
anisotropy spectrum.  Persi \etal\ (1995) estimate the
effect in a CDM model from hydrodynamic simulations and find
that this non-linear Doppler effect is small in comparison
to the second order Doppler (Vishniac) effect

Compton scattering off hot electrons also produces spectral
distortions which look like anisotropies to any experiment
confined to low frequency measurements.  As discussed in \S 2.1,
Compton scattering exchanges energy between the electrons
and photons to order $T_e/m_e$.  This upscatters photons
from the Rayleigh-Jeans to the Wien regime and leads
to a temperature distortion of $(\Delta T/T)_{RJ} = -2 y
\approx -2 \tau_c T_e /m_e \approx 10^{-5}-10^{-3}$ for
an individual cluster.  Again, the rms fluctuations would
be much smaller.  Much effort has been expended to
estimate the fluctuations caused by this {\it thermal} SZ
effect with varying results (Barbosa \etal\ 1995; Makino \& Suto 1993;
Cole \& Kaiser 1988).  Recently, empirical
modeling of clusters has predicted that the anisotropy at
arcminutes should be on the order of $(\Delta T/T)_{RJ} \simlt
10^{-7}$ (Ceballos \& Barcons 1994).
Moreover the signal is in large part due to bright
and easily identifiable clusters.  If such known clusters are removed
from the sample, the anisotropy drops to an entirely negligible level.

\titleb{4.2.5}{Inhomogeneous Reionization}

Finally, let us mention another variant of
the Vishniac-Sunyaev-Zel'dovich mechanism.  The optical depth
could vary due to inhomogeneities in the ionization fraction $x_e$
near last scattering.  Note that inhomogeneities well before
optical depth unity have little effect on the CMB due to the
cut off from the visibility function.
Details of this effect will of course depend on the exact model
for reionization, \ie\ the extent of the inhomogeneities.
Small inhomogeneities may be expected to behave as the Vishniac
effect; large inhomogeneities like the kinetic SZ effect.

\titlea{5} {Discussion}

It should now be clear that a great wealth of information
about cosmology and the model for structure formation lies
waiting to be observed in the CMB anisotropy spectrum.
The location of the first acoustic peak provides a robust
classical test for curvature in the universe if the fluctuations
are known to be either adiabatic or isocurvature.  The only caveat
here is that if the universe suffered early reionization
the acoustic effect may be so suppressed that it becomes unobservable.
In this case, the CMB will place a lower limit on the epoch
of reionization.
It is very likely that the curvature and/or thermal history
of the universe will be measured in the near future.

Barring early reionization, once
we obtain precise measurements of the
first peak and beyond, we should be able to extract much more
information.   The relative locations of the higher peaks can cleanly
separate adiabatic and isocurvature models independently of the
curvature and thermal history.
A precise measurement of the location
of the peaks can supply information on $\Lambda$ and $\Omega_0 h^2$.
The heights of the peaks yield even more information if some minimal
assumptions are made for the theory of structure formation.
Indeed, requiring consistency with large scale structure measurements
should eventually fix the model quite precisely.
The relative heights of the peaks give a robust
probe of the baryon content $\Omega_b h^2$.  In any given model,
the absolute heights yield a constraint on the matter content
of the universe $\Omega_0 h^2$ and perhaps the number, temperature
and mass of the neutrinos.
Combining these pieces of  information we can infer the Hubble constant.

It would seem that all the fundamental cosmological parameters are
encoded in the CMB anisotropy spectrum.  Yet, even discounting
the possibility of early reionization, how likely is it that
we will precisely measure them with the next generation of
experiments?
The results of Jungman \etal\ (1995) suggest that at least the
curvature
can be measured to better than $5\%$ accuracy with a full sky
map to half a degree resolution.
Precisely how much information can be extracted will depend in the
end on how severe foreground contamination from
synchotron radiation, free-free emission, interstellar cold
dust, and radio point sources  will be.
Many of the effects described here will require
$\Delta T/T$ to $10^{-6}$ accuracy to measure definitively.
With sufficient frequency coverage, there is hope of
distinguishing the background from the foreground signal
to employ
at least the clean patches of the sky for cosmology.

\bigskip
\noindent {\it Acknowledgments:}  I would like to thank E. Bunn, J. Silk,
D. Scott, N. Sugiyama, M. White as the ideas and results presented
here arise from our many collaborations.  In particular, numerical
results presented here are from the code of N. Sugiyama unless otherwise
stated.  M. Tegmark provided useful comments on a draft of these notes.
I would also like to thank the organizers of this school
E. Martinez-Gonzalez and
J.L Sanz for an enjoyable and productive meeting.

\begrefchapter{References}

\ref
Albrecht, A., Coulson, D., Ferreira, P. \& Magueijo, J. (1995):
	astroph-9505030

\ref
Barbosa, D., Bartlett, J.G., Blanchard, A. \& Oukbir, J. (1995):
	astroph-9511084

\ref
Bardeen, J.M. (1980): Phys. Rev. D {\bf 22} 1882

\ref
Bond, J.R. (1995): ``Theory and Observations of the CBR'',
	in Cosmology and Large Scale Structure, ed. by
	Schaeffer (Elsevier, Netherlands)

\ref
Bond, J.R. \& Efstathiou, G. (1984): ApJ Lett. {\bf 285} L45

\ref
Bond, J.R. \etal\ (1994): Phys. Rev. Lett. {\bf 72} 13

\ref
Cole, S. \& Kaiser, N. (1988): MNRAS {\bf 233} 637

\ref
Crittenden, R. \etal\ (1994): Phys. Rev. Lett. {\bf 71} 324

\ref
Crittenden, R. \& Turok, N. (1995): astroph-9505120

\ref
de Laix, A.A. \& Scherrer, R.J. (1995): astroph-9509075

\ref
Dodelson, S., Gates, E. \& Stebbins, A. (1995): astroph-9509147

\ref
Dodelson, S. \& Jubas, J. (1995): ApJ {\bf 439} 503

\ref
Doroshkevich, A.G., Zel'dovich, Ya.B. \& Sunyaev, R.A. (1978):
Sov. Astron. {\bf 22} 523

\ref
Durrer, R., Gangui, A. \& Sakellariadou (1995): astroph-9507035

\ref
Haehnelt, M.G. \& Tegmark, M. (1995): astroph-9507077

\ref
Hu, W. (1995): astroph-9508126

\ref
Hu, W., Scott, D. \& Silk, J. (1994): Phys. Rev. D. {\bf 49} 648

\ref
Hu, W., Scott, D., Sugiyama, N. \& White, M. (1995): Phys. Rev. D
 (in press, astroph-9505043)

\ref
Hu, W. \& Sugiyama, N. (1995a): ApJ {\bf 436} 456

\ref
Hu, W. \& Sugiyama, N. (1995b): Phys. Rev. D {\bf 51} 2599

\ref
Hu, W. \& Sugiyama, N. (1995c): astroph-9510117

\ref
Hu, W. \& White, M. (1995): A\&A (in press, astroph-9507060)

\ref
Jungman, G., Kamionkowski, M., Kosowsky, A. \& Spergel, D.N. (1995):
	astroph-9507080

\ref
Kaiser, N. (1983): MNRAS {\bf 202} 1169

\ref
Kaiser, N. (1984): ApJ {\bf 282} 374

\ref
Kodama, H. \& Sasaki, M. (1984): Prog. Theor. Phys. Supp. {\bf 78} 1

\ref
Kofman, L.A. \& Starobinskii, A.A. (1985): Sov. Astron. Lett.
{\bf 9} 643 (1985)

\ref
Kosowsky, A. (1995): astroph-9501045

\ref
Lidsey, J.E. \etal\ (1995): astroph-9508078

\ref
Ma, C.-P. \& Bertschinger, E. (1995): astroph-9506072

\ref
Magueijo, J., Albrecht, A., Coulson, D. \& Ferreira, P. (1995):
astroph-9511042

\ref
Makino, N. \& Suto, Y. (1993): ApJ {\bf 405} 1

\ref
Martinez-Gonzalez, E., Sanz, J.L. \& Silk, J. (1992): Phys. Rev. D {\bf 46}
4193

\ref
Mukhanov, V.F., Feldman, H.A., \& Brandenberger, R.H. (1992):
	Phys. Rep. {\bf 215} 203

\ref
Ostriker, J.P. \& Vishniac, E.T. (1986): ApJ {\bf 306} 51

\ref
Peebles, P.J.E. \& Yu, J.T. (1970): ApJ {\bf 162} 815

\ref
Pen, U.-L., Spergel, D.N. \& Turok, N. (1994): Phys. Rev. D {\bf 49} 692

\ref
Persi, F.M., Spergel, D.N., Cen, R. \& Ostriker, J.P. (1995):
ApJ {\bf 442} 1

\ref
Press, W. \& Vishniac, E.T. (1980): ApJ {\bf 239} 1

\ref
Rees, M.J. \& Sciama, D.N. (1968): Nature {\bf 519} 611

\ref
Sachs, R.K.  \& Wolfe, A.M. (1967): ApJ {\bf 147} 73

\ref
Scott, D., Silk, J. \& White, M. (1995): Science {\bf 268} 829

\ref
Seljak, U. (1994): ApJ Lett. {\bf 419} L47

\ref
Seljak, U. (1995a): astroph-9506048

\ref
Seljak, U. (1995b): astroph-9505109

\ref
Silk, J. (1968): ApJ Lett. {\bf 151} 459

\ref
Smoot, G. \etal\ (1992): ApJ Lett. {\bf 396} L1

\ref
Steinhardt, P.J. (1995): astroph-9502024

\ref
Sugiyama, N. \& Gouda, N. (1992): Prog. Theor. Phys. {\bf 88} 803

\ref
Sunyaev, R.A. \& Zel'dovich, Ya. B. (1972): Comm. Astrophys. Space
Phys. {\bf 4} 73

\ref
Sunyaev, R.A. \& Zel'dovich, Ya. B. (1980): MNRAS {\bf 190} 413

\ref
Tuluie, R. \& Laguna, P. (1995): ApJ Lett. {\bf 445} L73

\ref
Vishniac, E.T. (1987): ApJ {\bf 322} 597

\ref
Vittorio, N. \& Silk, J. (1984): ApJ Lett. {\bf 285} L39

\ref
Weinberg, S. (1972): Gravitation and Cosmology (Wiley, New York)

\ref
White, M., \& Scott, D. (1995): astroph-9508157

\ref
White, M., Scott, D. \& Silk, J. (1994): ARA\&A {\bf 32} 319 (1994)

\ref
Wilson, M.L. (1983): ApJ {\bf 273} 2

\ref
Wilson, M.L. \& Silk, J. (1981): ApJ {\bf 243} 14

\endref

\end